\documentclass[epj]{svjour}

\usepackage{graphics}
\usepackage{wrapft}
\usepackage{lscape}
\usepackage{amsmath}
\usepackage{bm}
\usepackage{here}
\usepackage{nidanfloat}
\newcommand{\Slash}[1]{{\ooalign{\hfil$#1$\hfil\crcr\raise.167ex\hbox{/}}}}
\newcommand{\ltsim}{\protect\raisebox{-0.5ex}{$\:\stackrel{\textstyle <}{\sim}\:$}}

\begin{document}

\title{Hyperon Puzzle, Hadron-Quark Crossover \\
 and Massive Neutron Stars
}

\author{
Kota Masuda \inst{1,2} \thanks{\emph{Present address:} masuda@nt.phys.s.u-tokyo.ac.jp}, Tetsuo Hatsuda\inst{2,3}  \and Tatsuyuki Takatsuka\inst{2} 
}                     

\institute{
Department of Physics, The University of Tokyo, Tokyo 113-0033, Japan \and Theoretical Research Division, Nishina Center, RIKEN, Wako 351-0198, Japan
\and Kavli IPMU (WPI), The University of Tokyo, Chiba 277-8583, Japan
}

\date{Received: date / Revised version: date}

\abstract{
Bulk properties of cold and hot neutron stars are studied 
on the basis of the hadron-quark crossover picture where a smooth transition 
from the hadronic phase to the quark phase takes place at finite baryon density. 
 By using a phenomenological equation of state (EOS) ``CRover" 
 which interpolates the two phases at around 3 times the nuclear matter density ($\rho_0$), 
  it is found that the cold NSs with the gravitational mass larger than
  $2M_{\odot}$ can be sustained. This is
  in sharp contrast to the  case of the first-order hadron-quark transition.
 The radii of the cold NSs with the CRover EOS 
 are in the narrow range $(12.5 \pm 0.5)$ km  which is insensitive to the NS masses.
 Due to the stiffening of the EOS induced by the hadron-quark crossover, 
 the central density of the NSs is at most 4 $\rho_0$ and 
  the  hyperon-mixing barely occurs inside the NS core.  
This constitutes a solution of the long-standing hyperon puzzle.
  The effect of color superconductivity (CSC) on the NS structures is also examined
  with the hadron-quark crossover.  For the typical strength of the diquark attraction,
 a slight softening of the EOS due to two-flavor CSC (2SC) takes place and   
 the maximum mass is reduced by about 0.2 $M_{\odot}$.
 The CRover EOS is generalized to the supernova matter at finite temperature 
  to describe the hot  NSs at birth.
  The hadron-quark crossover is found to decrease the central  temperature of the hot NSs
   under isentropic condition.
  The gravitational energy release and the 
  spin-up rate during the contraction from the hot NS to the cold NS are also estimated.  
} 

\authorrunning{Masuda, Hatsuda and Takatsuka}
\maketitle

\section{Introduction}
\label{intro}

\begin{figure*}[!t]
\begin{center}
\resizebox{0.9\textwidth}{!}{
  \includegraphics{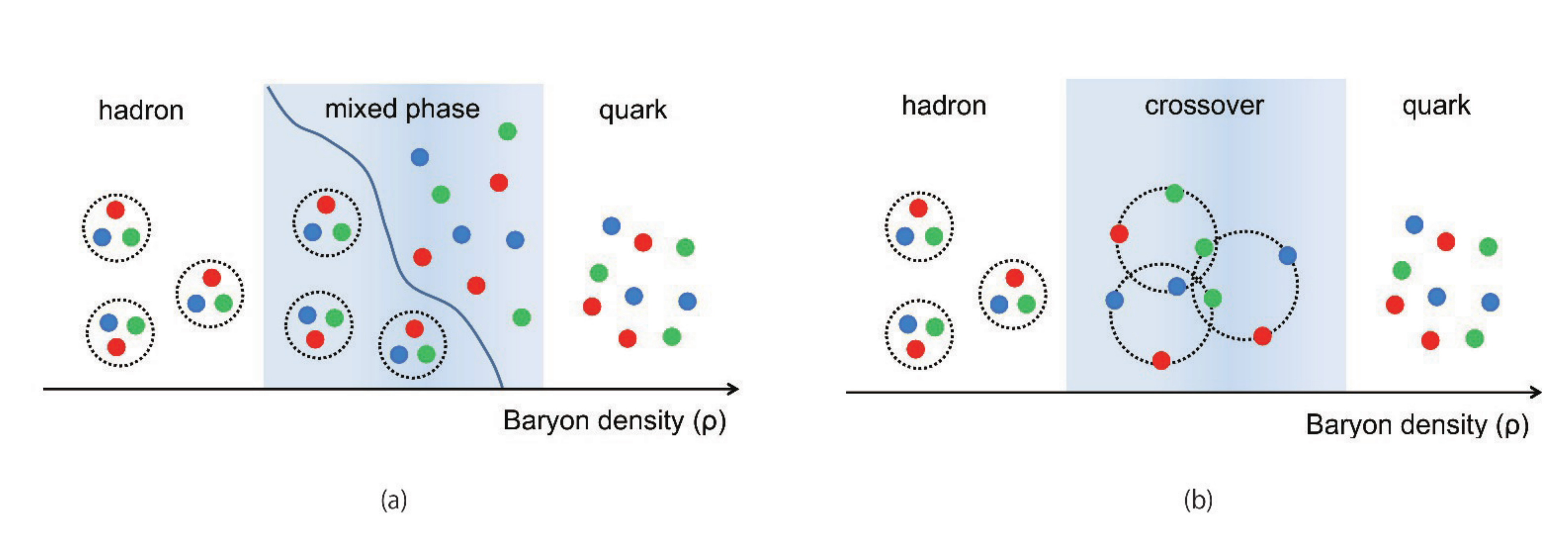}}
\caption{\footnotesize{Schematic pictures of the QCD phases as a function of 
 the baryon density ($\rho$) under the assumption of (a) the first-order transition
 and (b) the hadron-quark crossover.
 The mixed-phase region in (a) and the crossover region in (b) are
 indicated  by the shaded area.  }}
\label{crossover-1}
\end{center}
\end{figure*}

Strongly interacting matter described by quantum chromodynamics (QCD) is believed to
have a rich phase structure under the change of external parameters such as the temperature ($T$)
and the baryon chemical potential ($\mu$) \cite{fukushima-hatsuda}. 
At low $T$ and $\mu$, the system is in the hadronic phase where 
the dynamical breaking of chiral symmetry
and confinement of quarks and gluons take place.
At high $T$ and low $\mu$, the quark-gluon plasma with deconfined
quarks and gluons was predicted theoretically and 
is under active experimental studies 
by the relativistic heavy-ion collisions at RHIC and LHC \cite{HIC-review}. 
At low $T$ and high $\mu$, 
the superconducting quark matter with deconfined quarks  
is expected to appear, which is relevant to the central core of neutron stars \cite{Fukushima:2013rx}. 
 
The transition from the 
hadronic matter to the quark-gluon plasma at high $T$
has been  studied quantitatively by using the 
lattice QCD simulations \cite{Bazavov:2015rfa}.
On the other hand,  the transition from the hadronic matter to the 
quark matter  at high $\mu$ is not well understood 
partly due to the lack of reliable first-principle
theoretical methods; the Monte Carlo simulations in lattice QCD
are not suitable for $\mu/T \gg 1$ because of the fermion sign problem \cite{fukushima-hatsuda}. 
 
Under such circumstance,  any information from neutron stars (NSs), 
whose central cores may reach the baryon density relevant to the 
hadron-quark transition, is quite useful to understand the structure of high density matter. 
Among various observables for NSs \cite{Lattimer:2012nd}, 
the surface temperature ($T_s$)  the mass ($M$), the radius ($R$) and the magnetic field ($B$)
are particularly informative. Indeed, 
a massive NS (PSR J1614-2230) with $M= (1.97 \pm 0.04)M_{\odot}$ 
observed through the Shapiro delay technique \cite{Demorest:2010}  and 
another massive NS (PSR J0348+0432) with $M = (2.01 \pm 0.04)M_{\odot}$
\cite{Antoniadis} may give stringent constraints on the equation of state (EOS) of dense matter.

Historically, Gibbs phase equilibrium conditions
have been adopted to describe the transition
between the hadronic matter composed of 
point-like hadrons and the quark matter composed of  weakly interacting quarks.
However, in the transition region, 
neither the assumption of point-like hadrons nor that 
of weakly-interacting quarks are justified.
Indeed, there may arise a smooth crossover between the hadronic matter and 
the quark matter: Furthermore, the system in the crossover region would be
strongly interacting \cite{takatsuka-2011}.

Fig.~\ref{crossover-1} illustrates the
difference between (a) the first-order transition where
the phase separation between\\ hadrons and quarks takes place, and (b)
the crossover where the percolation of finite size hadrons takes place.
Such a percolation picture of hadrons
has been first discussed in Refs.\cite{baym,satz}, and 
later elaborated in the contexts of 
the  hadron-quark continuity \cite{Schafer-wilczek,Fukushima}
and the hadron-quark crossover \cite{Hatsuda:2006ps,Maeda}.

Recently, the present authors have 
shown that the \\hadron-quark crossover 
at around three times the normal nuclear matter density ($\rho_0=0.17$ fm$^{-3}$)
can lead to a stiffening of EOS and sustain the 2$M_{\odot}$ NSs
\cite{Masuda:2012kf,Masuda:2012ed,Masuda2015} 
in contrast to the case of hadron-quark  first-order transition. Also, it was shown that
such a stiffening due to hadron-quark crossover
can avoid the so-called  ``Hyperon Puzzle" as  discussed in \S 2.
(See also the related works \cite{Alvarez-Castillo:2013spa,Hell:2014xva,Kojo:2014rca}.) 
  
In this article, we discuss bulk properties of cold and hot NSs 
on the basis of the new EOS with the hadron-quark crossover (the ``CRover" EOS) 
introduced in \cite{Masuda:2012kf,Masuda:2012ed,Masuda2015}. 
In \S 2, we summarize the conventional hadronic EOS with and without hyperons.
In \S 3, we summarize detailed properties of hadronic EOS to be used
throughout the present study.
In \S 4, we summarize the quark EOS  based on the 
(2+1)-flavor Nambu--Jona-Lasinio (NJL) model at high density.
In \S 5,  We introduce a phenomenological approach to interpolate the 
hadronic matter and the quark matter. 
In \S 6,  we show the bulk properties of cold neutron stars using the CRover
EOS at $T=0$ (abbreviated as cold CRover EOS)
with and without color superconductivity.
In \S 7, we show the bulk properties of hot neutron stars at birth using CRover
EOS at $T\neq 0$ (abbreviated as hot CRover EOS).
\S 8 is devoted to summary and concluding remarks.

\section{Hyperon Puzzle}

\begin{figure*}[!t]
\begin{center}
\resizebox{0.9\textwidth}{!}{\includegraphics{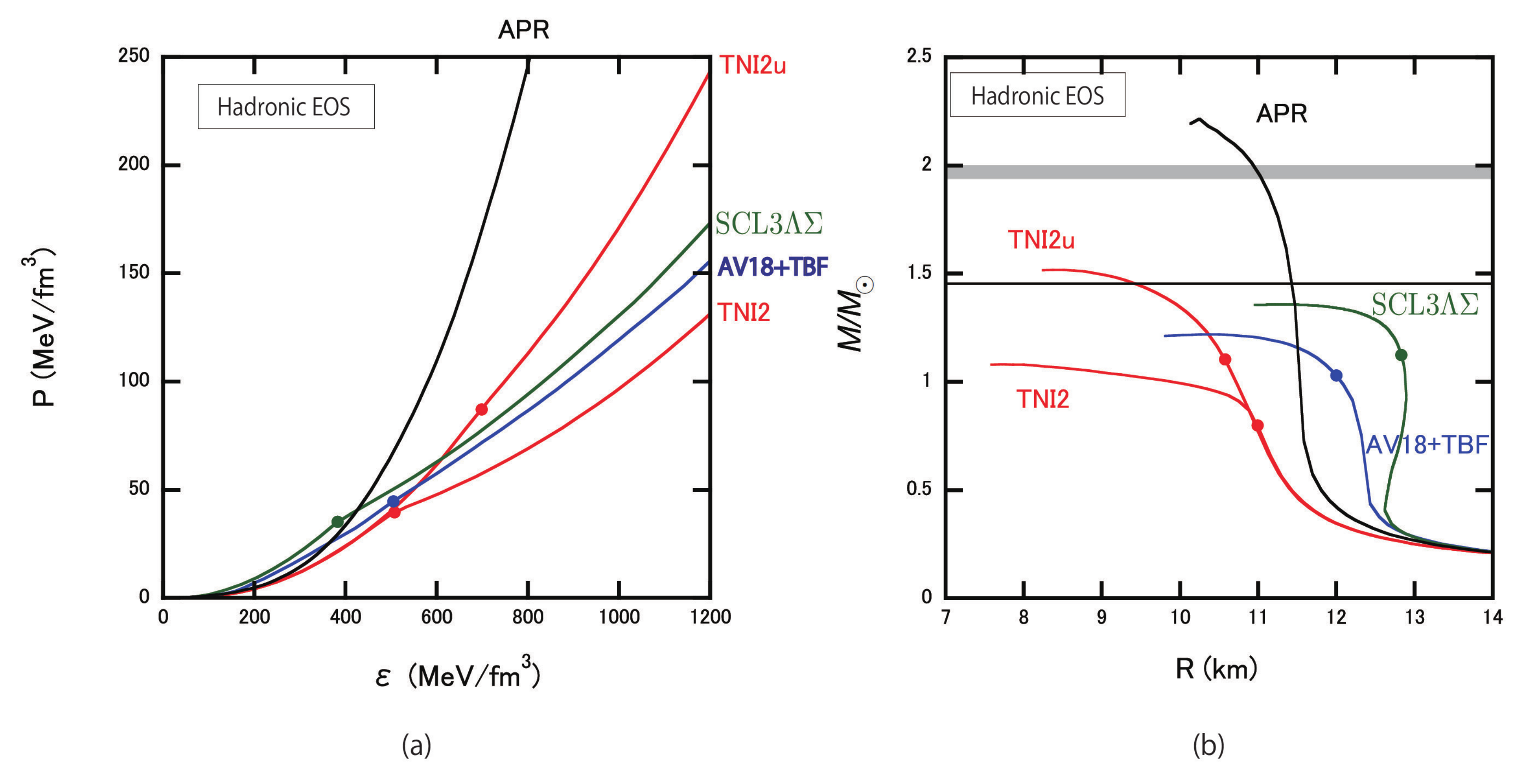}}
\caption {
(a) The hadronic equation of states with and without the hyperon mixing.
The black line denotes the EOS without hyperon, APR (AV18+$\delta v$+ UIX$^*$).
The red lines correspond to the EOS with hyperons;
TNI2 (only three-nucleon force with $\kappa=250$MeV) and
TNI2u (universal three-baryon force with $\kappa=250$MeV).
The blue line corresponds to   AV18+TBF+$\Lambda\Sigma$ (G-matrix with AV18 + 3-nucleon force + hyperons),
and the green line corresponds to  SCL3$\Lambda\Sigma$ (relativistic mean-field model  with  chiral SU(3) symmetry).
(b) $M-R$ relations for five EOSs considered in (a).
In both figures,  filled circles on each line show the   density where the hyperons start to mix.
Also,  the gray band denotes $M=(1.97\pm0.04)M_{\odot}$ for PSR J1614-2230 \cite{Demorest:2010}
and the solid horizontal line shows
$M  = 1.44 M_{\odot}$  corresponding to  PSR 1913+16.
Those figures are adapted and modified from \cite{Masuda:2012kf,Masuda:2012ed}. 
 }  
\label{fig:fig2}
\end{center}
\end{figure*}

Let us first consider what
would be the neutron star structure under the hadronic EOS with and without hyperons 
($Y$=$\Lambda$, $\Sigma^-$, $\Xi^-$). 
 Although there exist large uncertainties for the two-body $Y$-$N$ 
 interactions,\footnote{
 We note that it is important to have new data on hyperon interactions
 from the  $Y$-$N$ scattering and hypernuclei experiments at J-PARC \cite{Nagae,Tamura,Nakazawa}
 and also from the lattice QCD simulations at the physical quark masses \cite{HAL}.}
 various different models suggest that hyperons may appear
 at densities of several times $\rho_0$. 
 Those hyperons introduce significant softening of EOS and even the 
 well-established 1.4$M_{\odot}$ NSs may not be reproduced (see the reviews,
 \cite{takatsuka_2004,Vidana:2015faa} and the references therein.)
 The three-body force in the hyperon sector originally suggested in \cite{Nishizaki:2001}
 may or may not describe the 2$M_{\odot}$ NSs depending on its strength 
 \cite{takatsuka_2004,Takatsuka:2008,Yamamoto:2014jga,Lonardoni:2014bwa,Katayama:2015dga}. This is called the ``Hyperon Puzzle".
  
 To see the ``Hyperon Puzzle" more explicitly, 
 let us take four hadronic EOS with hyperons,
 TNI2u, TNI2, AV18+TBF+$\Lambda\Sigma$, and SCL3$\Lambda\Sigma$. 
 Here  TNI2u (TNI2) is the EOS based on the non-relativistic G-matrix approach
 with the incompressibility $\kappa=250$MeV and with (without) the hyperon three-body force.
 AV18+TBF+$\Lambda\Sigma$ \cite{Baldo:2000}  is also based on 
 the non-relativistic G-matrix approach with the AV18 nucleon-nucleon potential,
 the Urbana-type three-body nucleon potential and the Nijmegen soft-core nucleon-hyperon potential.
 SCL3$\Lambda\Sigma$ \cite{Tsubakihara:2010}
 is a relativistic mean-field model  with  chiral SU(3) symmetry. 
 As a typical nuclear EOS without the hyperons, we take APR \cite{Akmal:1998}.

 In Fig.\ref{fig:fig2}(a), we plot the hadronic EOS with hyperons
 (TNI2u, TNI2,  AV18+TBF+$\Lambda\Sigma$  and SCL3$\Lambda\Sigma$) together with APR.
 Filled circles on each line denote the
 density where the hyperon-mixing starts to occur.  
 One can see that (i) the mixture of hyperons softens the equation
 of state relative to APR, and (ii) onset of the hyperon-mixing is shifted to 
 higher density if we consider the three-body interaction in the hyperon sector.     
 In Fig.\ref{fig:fig2}(b), 
 the $M-R$ relations with these EOSs are plotted by the same color lines.
 The gray band shows  $M  = (1.97 \pm 0.04)M_{\odot}$  corresponding to
 PSR J1614-2230 \cite{Demorest:2010} and the solid horizontal line shows
 $M  = 1.44 M_{\odot}$  corresponding to  PSR 1913+16.
 In Table \ref{property of H-EOS}, 
 we summarize the nuclear incompressibility $\kappa$, 
 the threshold density for $\Lambda$ and $\Sigma^-$, 
 the maximum mass $M_{\rm max}$, the radius and the central density $\rho_c$ for these hadronic EOSs 
 with hyperons. EOSs become soft drastically due to the emergence of hyperons.

\begin{table}[!b]
\caption{\footnotesize{Properties of various hadronic EOSs with hyperons;
  TNI2, TNI2u, AV18+TBF+$\Lambda \Sigma$ and SCL$3\Lambda \Sigma$. 
  $\kappa$ is the nuclear incompressibility and 
  $\rho_{\rm th}$ is the threshold density of 
  hyperon-mixing with $\rho_0$ being the normal nuclear density. 
  $R$ and $\rho_c$ denote
the radius and central density of the maximum mass ($M_{\rm max}$) NS, respectively.
 The numbers in the parentheses are those without hyperons.
 ``$\ast$"s indicate that the numbers are read from the figures in \cite{Baldo:2000}.
  } }
 \label{property of H-EOS}
\begin{center}
  \begin{tabular}[c]{c|c|c|c|c}  \hline \hline
  EOS & TNI2  & TNI2u & AV18+TBF & SCL$3\Lambda \Sigma$  \\ \hline
$\kappa$ (MeV) & 250 & 250  & 192 & 211      \\ 
$\rho_{\rm th}(\Lambda)/\rho_0$  & 2.95 & 4.01  & 2.8$^{\ast}$ & 2.24      \\ 
$\rho_{\rm th}(\Sigma^-)/\rho_0$ & 2.83 & 4.06  & 1.8$^{\ast}$  & 2.24     \\ \hline
$M_{\rm{max}}/M_{\odot}$  & 1.08 & 1.52  & 1.22 & 1.36      \\
                          &(1.62)&  & (2.00)      &  (1.65)         \\
R(km)                     & 7.70 & 8.43  & 10.46 & 11.42   \\
                          &(8.64)& & (10.54)     & (10.79)        \\
$\rho_c/\rho_0$           & 16.10& 11.06 &7.35 & 6.09      \\
                          &(9.97)& & (6.53)      & (6.85)          \\
\hline \hline
\end{tabular} 
\end{center}
\end{table}

\section{Hadronic EOS with Hyperons}

\begin{figure*}[!t]
\begin{center}
\resizebox{0.9\textwidth}{!}{
  \includegraphics{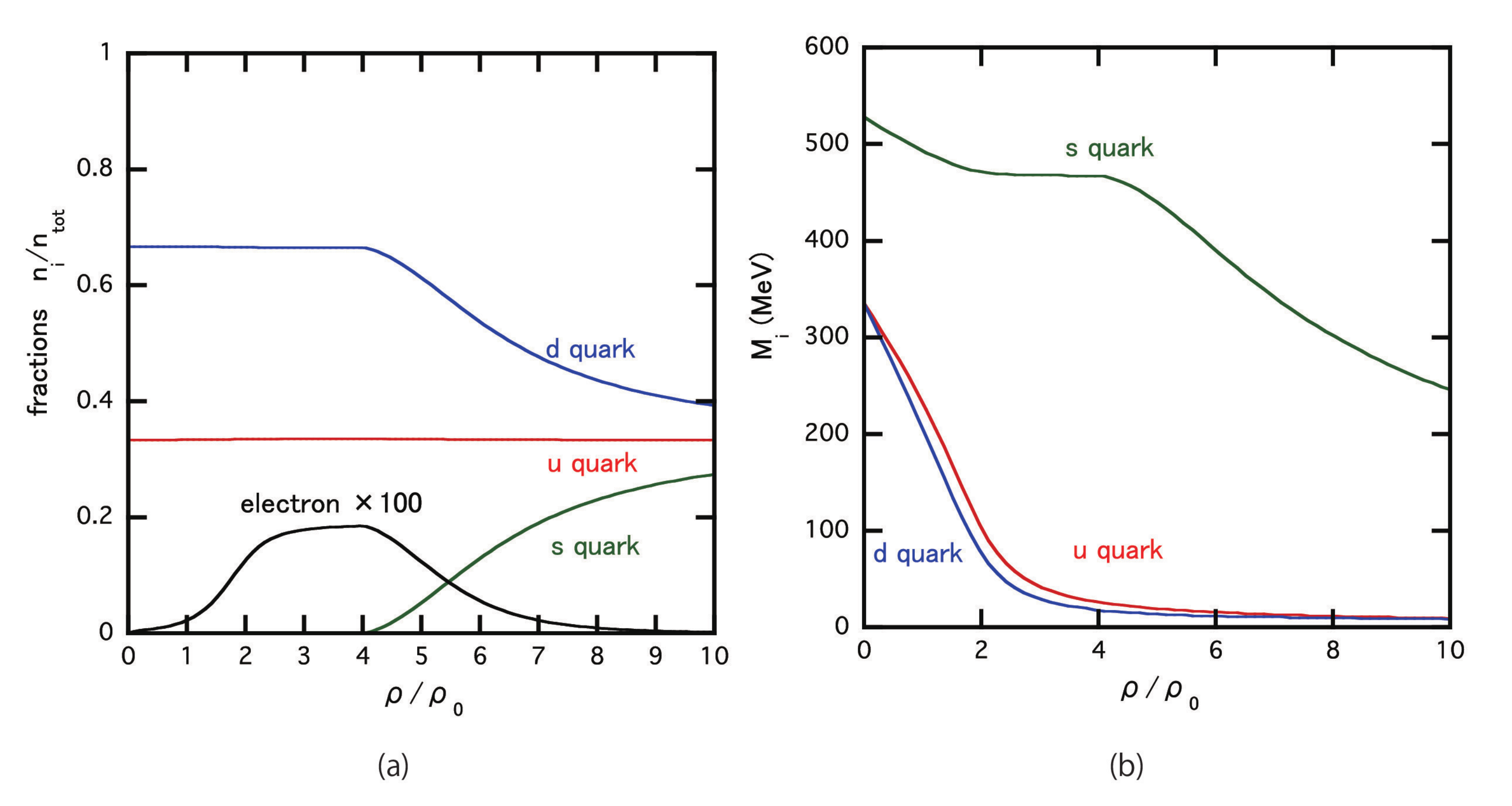}
  }
\caption{\footnotesize{
(a)
The number fractions ($ n_{u,d,s,e}/n_{\rm tot}$ with
$n_{\rm tot}=n_u+n_d+n_s=3 \rho$) as a function of the baryon density $\rho$.
Red, blue and green lines correspond to  u, d and s quark, respectively.
The black line corresponds to the electron number fraction $\times$ $100$. 
Muon does not appear due to the emergence of s quarks.
(b)
The constituent quark masses ($M_i$) as a function of $\rho$.
Colors on each line are the same with those in (a). 
These figures are adapted from \cite{Masuda:2012ed}.
}}
\label{quark-eos}
\end{center}
\end{figure*}

Since we will take TNI2u and TNI2 
in the following analyses, we summarize here how these
hadronic EOS with hyperons  are obtained
\footnote{``TNI" implies that the Three-Nucleon Interaction is taken into account, 
``2" implies  $\kappa=250$ MeV, and ``u" implies that the three-body interaction
is introduced universally in the octet baryon sector.}  \cite{Nishizaki:2001,Takatsuka:2005bp}:
\begin{enumerate}
\item
Effective two-baryon potentials ${\tilde V}_{BB'}$ ($B= n$, $p$, $\Lambda$, 
$\Sigma^-$) are constructed on the
 basis of the G-matrix formalism to take into account their density-dependence.
\item
  A phenomenological thee-nucleon interaction 
  expressed in a form of two-body potential ${\tilde U}_{NN'}$  \cite{Friedman:1981qw}
  is introduced to reproduce the saturation of symmetric  nuclear matter
  (the saturation density $\rho_0=0.17$ fm$^{-3}$ and the binding 
energy $E_0=-16$ MeV)  and the incompressibility $\kappa=250$ MeV compatible with experiments.
\item
  Universal three-body repulsion is assumed 
  for the hyperons in TNI2u through the replacement,
  ${\tilde U}_{NN'} \rightarrow {\tilde U}_{BB'}$, which is 
  necessary to sustain 1.4$M_{\odot}$ even with hyperons, while
  the three-body repulsion is introduced only in the nucleon sector in TNI2.
\item
   By using ${\tilde V}_{BB'}+{\tilde U}_{BB'}$,
   we calculate the  particle composition $y_i$ ($i =n$, $p$, $\Lambda$, $\Sigma^-$, $e^-$ and $\mu^-$)
   under charge neutrality and $\beta$-equilibrium to obtain the hadronic EOS
   as a function of total baryon density $\rho$ at $T=0$.
 \end{enumerate}
    As shown in Fig.\ref{fig:fig2}(a), TNI2u EOS is 
    moderately stiff even with hyperon-mixing, but the corresponding maximum mass of NS is 1.5$M_{\odot}$,
    so that it is  not sufficient to reproduce  2$M_{\odot}$ NSs as shown in Fig.\ref{fig:fig2}(b).

\section{(2+1)-flavor Quark EOS with color superconductivity}
\label{sec:NJL}

\begin{figure*}[!t]
\begin{center} 
\resizebox{0.9\textwidth}{!}{
  \includegraphics{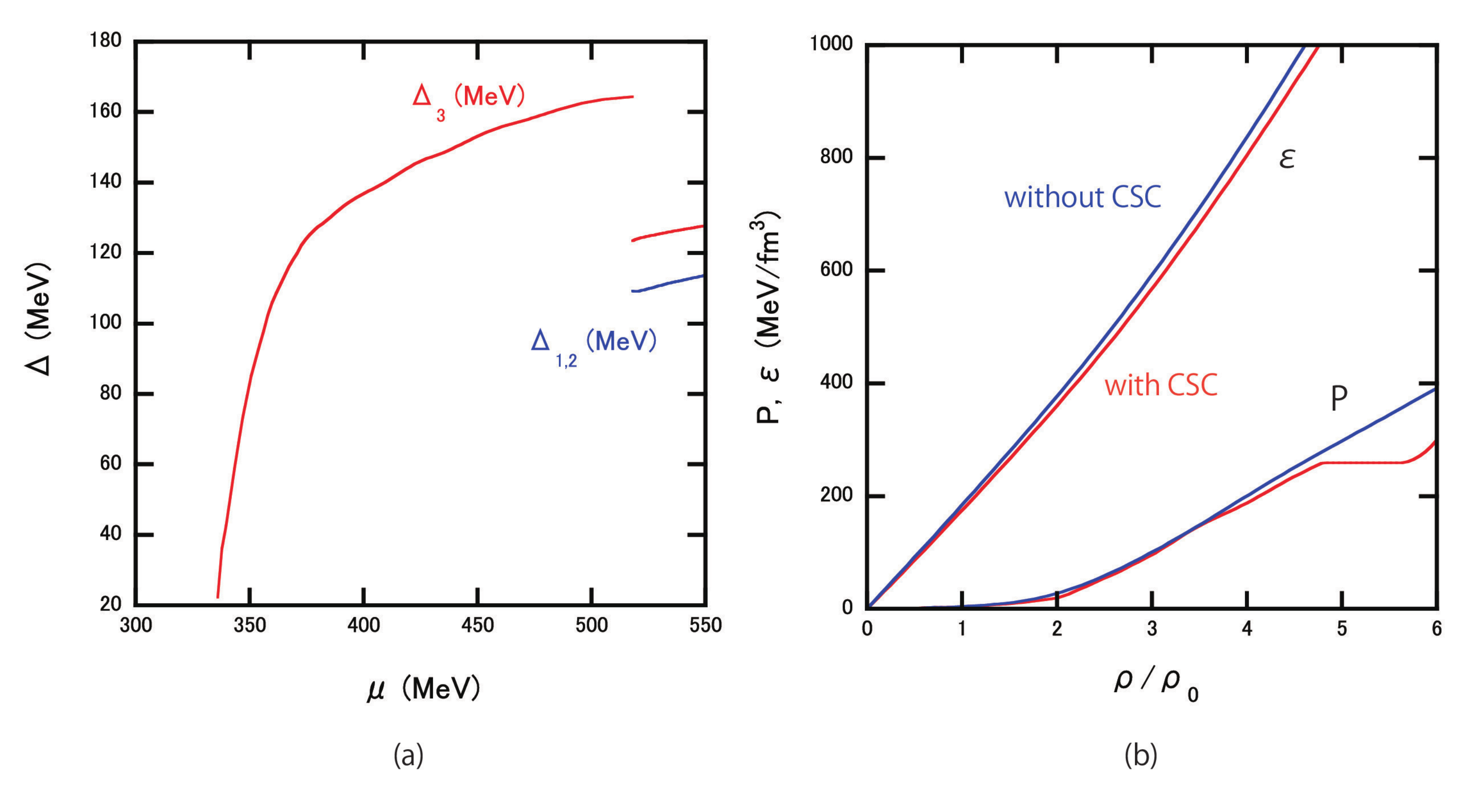}
  }
\caption{\footnotesize{
(a)
The gap parameters $\Delta_i$ ($i=1,2,3$) as a function of the quark chemical potential $\mu$.
Solid red line: $\Delta_3$
corresponding to the pairing between $u$ quark and $d$ quark. 
Solid blue line: $\Delta_1$ and $\Delta_2$
corresponding to the $ds$ pairing and $su$ pairing, respectively. 
(b)
The energy density ($\varepsilon$) and pressure ($P$) as a function of baryon number density $\rho$.
The red lines show the quark EOSs with diquark condensate.
The blue lines  show the quark EOSs without diquark condensate. 
}}
\label{gap parameter}
\end{center}
\end{figure*}

At high baryon density relevant to the central core of the NSs, 
baryons are started to overlap with each other and quark degrees of freedom may arise 
with $\mu \sim 400-500$ MeV.
However, at such chemical potentials, the QCD running coupling is still large
and the quark matter would be   strongly interacting.
 Analogous strongly interacting matter at finite $T$ 
 was originally  discussed 
 theoretically in \cite{Linde:1980ts,Hatsuda:1985eb,DeTar:1985kx}
 and was later studied experimentally  in the relativistic heavy-ion collisions at RHIC and LHC  \cite{HIC-review}.
 
 Under the situation that the Monte Carlo simulations in lattice QCD are not applicable to 
 $\mu/T \gg 1$ due to the sign problem,
 we adopt the (2+1)-flavor Nambu$-$Jona-Lasinio (NJL) model 
 which captures the essential properties of QCD such as the dynamical breaking of 
 chiral symmetry and its restoration at finite $T$ and $\mu$ \cite{Hatsuda-Kunihiro,Buballa}.
 The model Lagrangian we consider reads
\begin{eqnarray*}
{\mathcal L}_{\rm NJL}
&=&\overline{q}(i \Slash \partial -m)  q+\frac{1}{2}G_{_S}\sum_{a=0}^{8}[(\overline{q}\lambda^a q)^2+(\overline{q}i\gamma_5\lambda^a q)^2] \nonumber \\
&&- G_{_D}[\mathrm{det}\overline{q}(1+\gamma_5) q+ {\rm h.c.}]-\frac{1}{2}g_{_V}(\overline{q}\gamma^{\mu}q)^2 \nonumber \\
&&+\frac{H}{2}\sum_{I,A=2,5,7}(\bar{q}i\gamma_5 \lambda^{I} \tau^A C\bar{q}^T)(q^T C i\gamma_5 \lambda^{I} \tau^A q) \nonumber \\
&& +
\frac{G_{_D}'}{8}\sum_{i,j,k=1,2,3}
[(q \tilde{\lambda}_i \tilde{\tau}_k (1+\gamma_5)C\bar{q}^T) \nonumber \\
&&( \bar{q}\tilde {\lambda}_j \tilde{ \tau}_k(1+\gamma_5)Cq)(\bar{q}_i (1+\gamma_5) q_j)+\mathrm{h.c.}]
\label{eq-1}
\end{eqnarray*}
where the quark field $q_i^{\alpha}$ has three colors labeled by $\alpha$ and three flavors labeled by
$i$ with the current quark masses $m_i$.
The term proportional to $G_{_S}$ is a $U(3)_L \times U(3)_R$ symmetric
four-fermi interaction where  $\lambda^a$ are the Gell-Mann matrices in flavor space with $\lambda^0=\sqrt{2/3}\ {\rm I}$. 
The  term proportional to $G_{_D}$ is called as the Kobayashi$-$Maskawa$-$'t Hooft (KMT) 
coupling which breaks $U(1)_A$ symmetry explicitly \cite{Kobayashi:1970ji,'tHooft:1976fv}.
The term proportional to $g_{_V}(\ge 0)$ 
gives a universal repulsive force among different flavors.
The term proportional to $H$ gives a diquark condensation 
with color anti-triplet, flavor anti-triplet and $J^P=0^+$ channel
where $C=i\gamma_2\gamma_0$ is the charge conjugation matrix 
and $\tau^a$ are the Gell-Mann matrices in color space with $\tau^0=\sqrt{2/3}\ {\rm I}$.
The term proportional to $G_{_D}'$ is obtained by the Fierz transformation of the KMT term and
gives a coupling between the chiral condensate and the diquark condensate. Here we have introduced a notation, 
$\tilde {\lambda}_{1,2,3} \equiv \lambda_{7,5,2}$ and 
$\tilde {\tau}_{1,2,3} \equiv \tau_{7,5,2}$.

In the mean field approximation,
the constituent quark masses $M_i$ and the gap parameters $\Delta_i$ are generated dynamically 
through the NJL interactions,
\begin{eqnarray}  
M_i=m_i-2G_{_S} \sigma_i +2G_{_D} \sigma_j \sigma_k+\frac{G_{_D}'}{4}|s_i|^2,
\end{eqnarray} 
\begin{eqnarray}
\Delta_i=- \left( H- \frac{G_{_D}'}{2}\sigma_i \right)s_i,
\end{eqnarray}
where $\sigma_i = \langle \bar{q}_iq_i \rangle$
is the chiral condensate in each flavor, 
$s_i = \langle \bar{q}^T C\gamma_5 \tilde{\lambda}_i \tilde{\tau}_i q \rangle$
is the diquark condensate in each color and flavor with
 $(i,j,k)$ corresponding to the cyclic permutation of $u,d$ and $s$. 
The thermodynamic potential $\Omega$ is related to the pressure
as $\Omega=-T {\rm log} Z = -P V$ with $P$ given by
\begin{eqnarray}
P(T,\mu_{u,d,s})&&=
\frac{T}{2}
\sum_{\ell} 
\int \frac{d^3p}{(2\pi)^3}\mathrm{Trln}
\left(\frac{S^{-1}(i\nu_{\ell},{\bf{p}})}{T}\right)
\nonumber \\
&&-G_{_S}\sum_{i}\sigma_{i}^2
-4G_{_D}\sigma_{u}\sigma_{d} \sigma_{s}
+\frac{g_{_V}}{2}\left(\sum_i n_i \right)^2 \nonumber \\
&&-\sum_{i=1,2,3} \frac{1}{2}(H-G_{_D}')|s_i|^2.
\label{eq:NJL-pressure}
\end{eqnarray}
Here $i\nu_{\ell} =(2\ell +1)\pi T$ is the Matsubara frequency, 
$n_i = \langle q^{\dagger}_iq_i \rangle$ 
is the quark number density in each flavor,
and  $S$ is the quark propagator in the Nambu-Gor'kov representation, 
\begin{eqnarray}
\left[ S^{-1}\right]_{\alpha \beta}^{ij} =
\left(
    \begin{array}{cc}
      [G_0^+]^{-1} & \sum_{i=1,2,3}\Delta_i \gamma_5 \tilde{\lambda}_i \tilde{\tau}_i \\
      -\sum_{i=1,2,3}\Delta^{\ast}_i \gamma_5 \tilde{\lambda}_i \tilde{\tau}_i  & [G_0^-]^{-1}
    \end{array}
  \right) \nonumber
\end{eqnarray}
where
\begin{eqnarray}
[G_0^{\pm}]^{-1}=
\Slash p-\hat{M} \pm \gamma_0
\hat{\mu}^{\rm{eff}}.
\label{eq:S-prop}
\end{eqnarray}
Here,
$\hat{M}$ is a unit matrix in color space and a diagonal matrix in flavor space, ${\rm diag}(M_u,M_d,M_s)$.
The effective chemical potential matrix  $\hat{\mu}^{\rm{eff}}$ is defined from
\begin{eqnarray}
\hat{\mu}^{\rm{eff}}
\equiv \hat{\mu}-g_{_V}\sum_j n_j
\end{eqnarray}
where each component of $\hat{\mu}$ is given by
\begin{eqnarray}
\mu_{\alpha \beta}^{ij}=(\mu \delta^{ij}+\mu_Q Q^{ij})\delta_{\alpha \beta}
+(\mu_3 (\tau_3)_{\alpha \beta}+\mu_8 (\tau_{8})_{\alpha \beta})\delta^{ij}.
\nonumber
\end{eqnarray} 

\begin{figure*}[!t]
\begin{center}
\resizebox{0.9\textwidth}{!}{
  \includegraphics{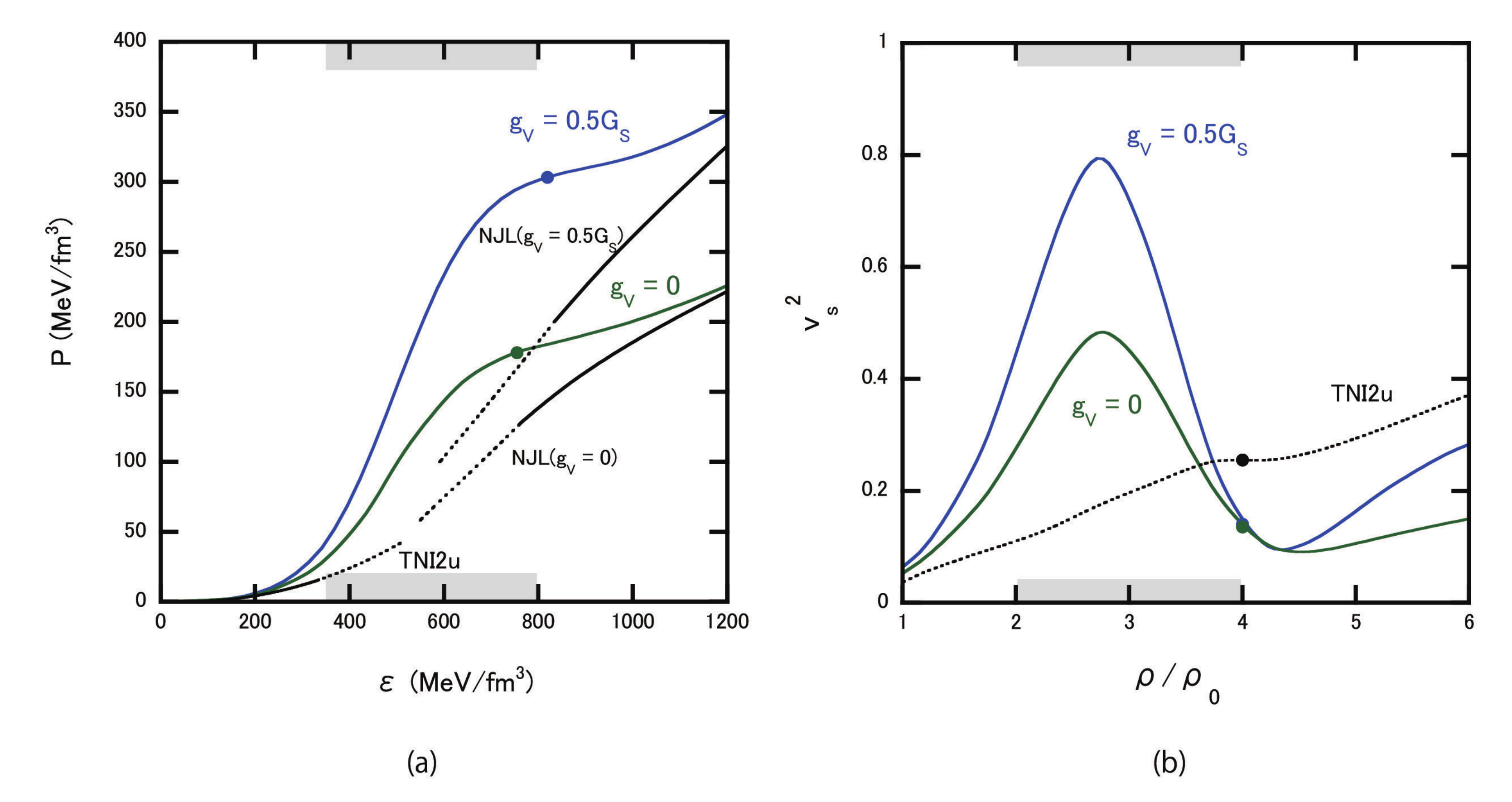}
  }
\caption{\footnotesize{
(a)
The relation between the interpolated energy density and the resultant pressure
 for  $g_{_V}=0$ (green) and $g_{_V}=0.5 G_{_S}$ (blue) without  CSC, $H=G_D'=0$.
 The crossover window is  $(2-4)\rho_0$. 
The filled circle denotes the threshold density of the strangeness.
(b)
Sound velocity squared $v_s^2$ as a function of baryon density $\rho$.
Solid lines show $v_s-\rho$ obtained from the
interpolated EOS  with $g_{_V}/G_{_S}=0, 0.5$, while the dotted line
corresponds to the TNI2u hadronic EOS. 
The filled circles denote the points beyond which strangeness starts to appear.
Those figures are adapted and modified from \cite{Masuda:2012ed}. 
}}
\label{interpolate-e-tni2u-sound}
\end{center}
\end{figure*}

\indent
There are nine independent 
parameters in the (2+1)-flavor NJL model; the UV cutoff, $\Lambda$, the 
coupling constants, $G_{_S},G_{_D}, g_{_V},H$ and $G_{_D}'$ and the current quark masses,
 $m_{u,d}$ and $m_{s}$. 
Five parameters except for $g_{_V},H$ and $G_{_D}'$
have been determined from hadron phenomenology in the vacuum.
In this article,
we adopt the HK parameter set \cite{Hatsuda-Kunihiro} (results for other parameter sets, 
see \cite{Masuda:2012ed});
\begin{eqnarray}
&&\Lambda=631.4 {\rm MeV},\ G_{_S}\Lambda^2=3.67,\ G_{_D}\Lambda^5=9.29, \nonumber \\
&&m_{u,d}=5.5 {\rm MeV},\ m_s=135.7{\rm MeV}.
\end{eqnarray} 
For $g_{_V}$, we change its magnitude in the following range \cite{Bratovic:2012qs,Lourenco},
\begin{eqnarray} 
  0 \le \frac{g_{_V}}{G_{_S}} \le 0.5 ,
  \label{eq:GV}
\end{eqnarray}

The parameters $H$ and $G_{_D}'$ are  chosen to be 
 $H = G_{_S}$ and $G_{D}'=G_{_D}$ as characteristic values.  
 (Corresponding values from the  Fierz transformation are $H=\frac{3}{4} G_{_S}$
 and $G_{D}'=G_{_D}$. 
 For extensive analyses with other choice of 
 parameters in the diquark channels, see \cite{Hatsuda-csc,Yamamoto-csc,Abuki-csc}.)

The EOS of quark matter with strangeness is obtained from the above model under two conditions: 
(i) the charge neutrality among quarks and leptons, 
$\frac{2}{3}n_u-\frac{1}{3}n_d-\frac{1}{3}n_s-n_e - n_{\mu}=0$,
(ii) the color neutrality among quarks,
 $n_r=n_g=n_b$,
and (iii) the $\beta$-equilibrium among quarks and leptons, 
$\mu_d=\mu_s=\mu_u+\mu_e$ and $\mu_e=\mu_{\mu}$. 
The ground state of the system is obtained by searching the maximum of the pressure in Eq. (\ref{eq:NJL-pressure}) with the conditions,
\begin{eqnarray}
\frac{\partial P}{\partial \sigma_{u,d,s}}=
\frac{\partial P}{\partial \Delta_{1,2,3}}=
\frac{\partial P}{\partial \mu_{3,8}}=0.
\label{eq:coupled-gapE}
\end{eqnarray}

Let us first discuss a composition  of the $\beta$-equilibrated quark matter 
 at $T=0$ without  color superconductivity ($H= G_{_D}'=0$).
In Fig.\ref{quark-eos} (a), the number fractions ($ n_{u,d,s,e}/n_{\rm tot}$ with
$n_{\rm tot}=n_u+n_d+n_s=3 \rho$) as a function of the baryon density $\rho$ are plotted. 
In Fig.\ref{quark-eos} (b), 
the constituent quark masses ($M_i$) as a function of $\rho$ are plotted.
These figures do not depend on the magnitude of the vector type interaction $g_{_V}$
because the flavor-independent $g_{_V}$-type interaction leads to a
pressure in Eq.(\ref{eq:NJL-pressure}) depending only on  $\mu_{\alpha,a}^{\rm eff}$.

At low densities, the $s$ quark appears only above $\rho_{\rm th}\simeq 4 \rho_0$ due to its heaviness 
as can be seen from Fig.\ref{quark-eos} (a): Here
 $\rho_{\rm th}$ is determined by the condition, $\mu_s(\rho_{\rm th}) = M_s(\rho_{\rm th})$.
The dynamical masses of $u$ and $d$ quarks decrease rapidly below $\rho_{\rm th}$ due to
partial restoration of chiral symmetry, while the $s$ quark is affected only a little 
through the KMT interaction as seen from Fig.\ref{quark-eos} (b).
  Once the $s$-quark whose electric charge is negative starts to appear above $\rho_{\rm th}$,
the number of electrons decreases to satisfy the charge neutrality.
Since the electron chemical potential does not exceed the muon mass $106$MeV,
the muons do not appear even at high density. 
In the high density limit, the system approaches to the flavor symmetric $u,d,s$ matter
without leptons. 

\begin{figure*}[!t]
\begin{center}
\resizebox{0.9\textwidth}{!}{
  \includegraphics{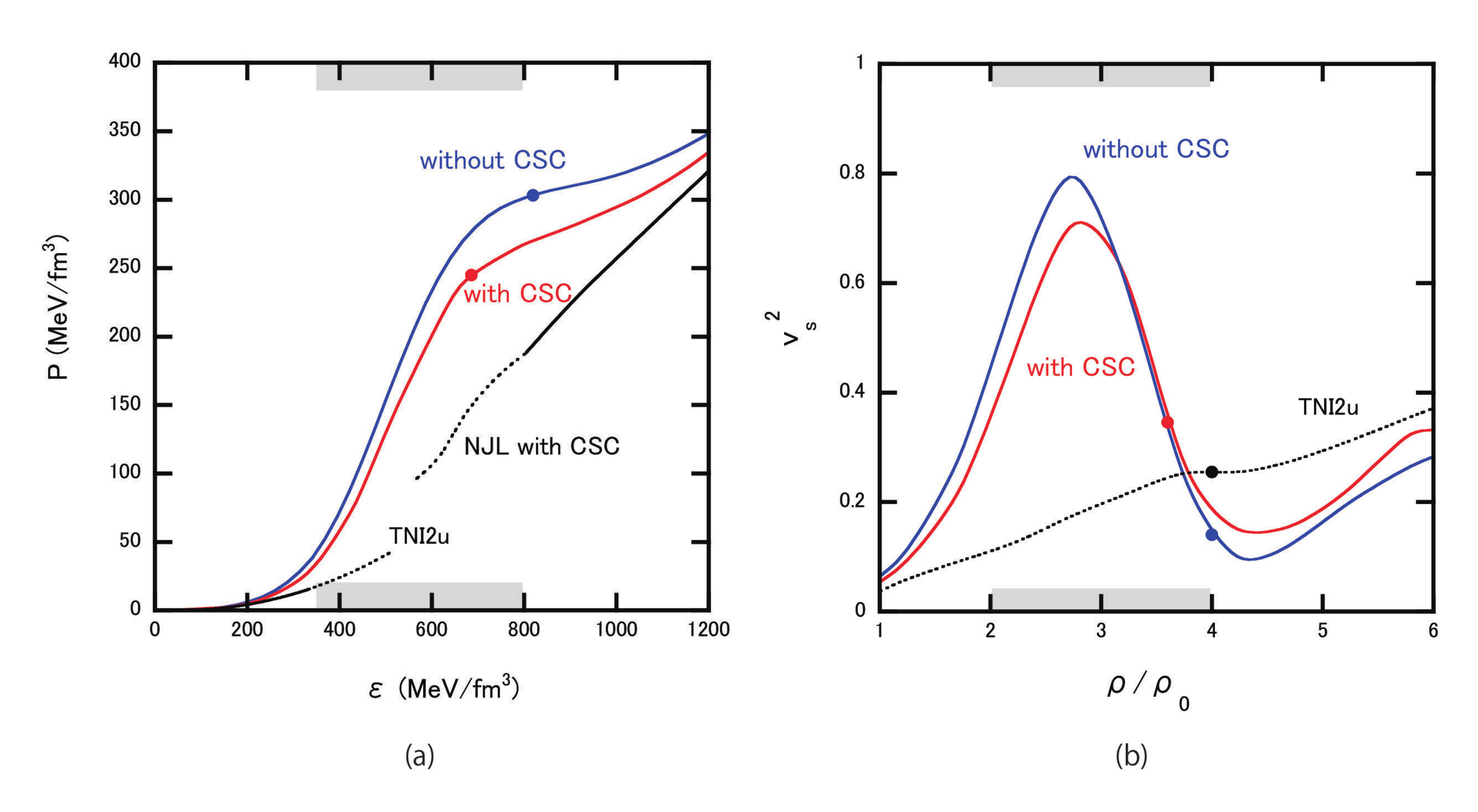}
  }
\caption{\footnotesize{
(a,b)
Comparison between 
the CRover EOS and the sound velocity squared without CSC 
($H=G_D'=0$) and those with CSC ($H=G_S$ and $G_D'=G_D$) for $g_{_V}/G_{_S}=0.5$.
}}
\label{csc-sound}
\end{center}
\end{figure*}

Once  the interactions in the diquark channels ($H$ and $G_{_D}'$) are switched on,
the color superconductivity (CSC) develops as shown in Fig. \ref{gap parameter} (a) where
the gap parameters $\Delta_i$ as a function of the quark chemical potential $\mu$ with $g_{_V}/G_{_S}=0.5$ are plotted.
The red line corresponds to the $ud$ pairing,
and blue line corresponds to the $ds$ or $su$ pairings.
With $H/G_{_S}=1$, two-color superconductivity (2SC) appears as soon as the 
 baryon density of the quark matter becomes finite at $\mu = 335 $ MeV. Then the
first-order transition from the 2SC to the color-flavor-locking (CFL) phase 
takes place at around $\mu =520 $MeV.
The diquark condensates affect the behavior of  the number fractions ($ n_{i,e}/n_{\rm tot}$
and the constituent quark masses ($M_i$) through the coupled equations, Eq.(\ref{eq:coupled-gapE}):
Those are taken into account into the following calculations with CSC.

Fig. \ref{gap parameter} (b) shows the thermodynamic quantities ($P$ and $\varepsilon$)
as a function of $\rho$ in the (2+1)-flavor NJL model.
The red (blue) lines correspond to the case with (without) CSC.
In terms of the baryon density,
2SC (CFL) appears for  $2 \rho_0 < \rho <  5 \rho_0$ ($\rho > 5 \rho_0)$
 in the present parameter set. The plateau of the red line ($P$ with CSC)
 reflects the fact that there is a  first-order transition from 2SC to CFL.  
As we will see later, baryon density relevant to neutron stars with 
the hadron-quark crossover  is below $5\rho_0$. Therefore, the CFL phase 
barely appears in the central core of the NSs in the present model.

\section{CRover: New EOS with Hadron-Quark Crossover}

We now introduce the following phenomenological interpolation of the energy per baryon 
$\hat{E}\equiv E/N_{_B}$ at $T=0$ \cite{Masuda:2012kf,Masuda:2012ed} 
\begin{eqnarray}
\hat{E}(\rho)=\hat{E}_{\rm H}(\rho) w_-(\rho)+\hat{E}_{\rm Q}(\rho)  w_+ (\rho), 
\end{eqnarray}
where $\hat{E}_{\rm H}$ and $\hat{E}_{\rm Q}$ represent energy per baryon in the hadronic matter with 
leptons and that in the quark matter with leptons, respectively.
$w_{-}$ and $w_+=1-w_-$ are the  weight functions
\begin{eqnarray}
w_{\pm}(\rho) \equiv
 \frac{1}{2} \left( 1\pm \mathrm{tanh}\left( \frac{\rho-\bar{\rho}}{\Gamma} \right) \right),
\end{eqnarray}
where  $\bar{\rho}$ and $\Gamma$ are the phenomenological parameters which 
characterize the averaged crossover density and the width of the crossover window, respectively.
Similar weight function has been previously used to interpolate the hadronic phase and the quark-gluon plasma 
 at finite temperature  \cite{Blaizot-1,Asakawa-Hatsuda}.
The window $ \bar{\rho} - \Gamma \ltsim \rho \ltsim  \bar{\rho} + \Gamma$
characterizes the crossover region in which 
both hadrons and quarks are strongly interacting, so that neither pure hadronic EOS nor pure quark EOS are reliable.

The other observables can be obtained by using the thermodynamic relations from
$\hat{E}(\rho)=\varepsilon / \rho$;
\begin{eqnarray}
&&P= \rho^2 \frac{\partial \hat{E}}{\partial \rho}, \ \ 
\mu = \frac{\partial \varepsilon}{\partial \rho}, \ \ 
 K = \rho \frac{\partial P}{\partial \rho} ,\nonumber \\
&& v_s^2 (\rho)=  \frac{\partial P}{\partial \varepsilon} = \frac{K}{\varepsilon + P}.
\label{eq:P-mu-K-v} 
\end{eqnarray}
Here $K$ is the bulk modulus which must be positive semi-definite for thermodynamic stability.
 Also, $v_s$ is the sound velocity which is a characteristic measure of the 
stiffness of the EOS.

\begin{figure*}[!t]
  \begin{center}
  \resizebox{0.9\textwidth}{!}{
  \includegraphics{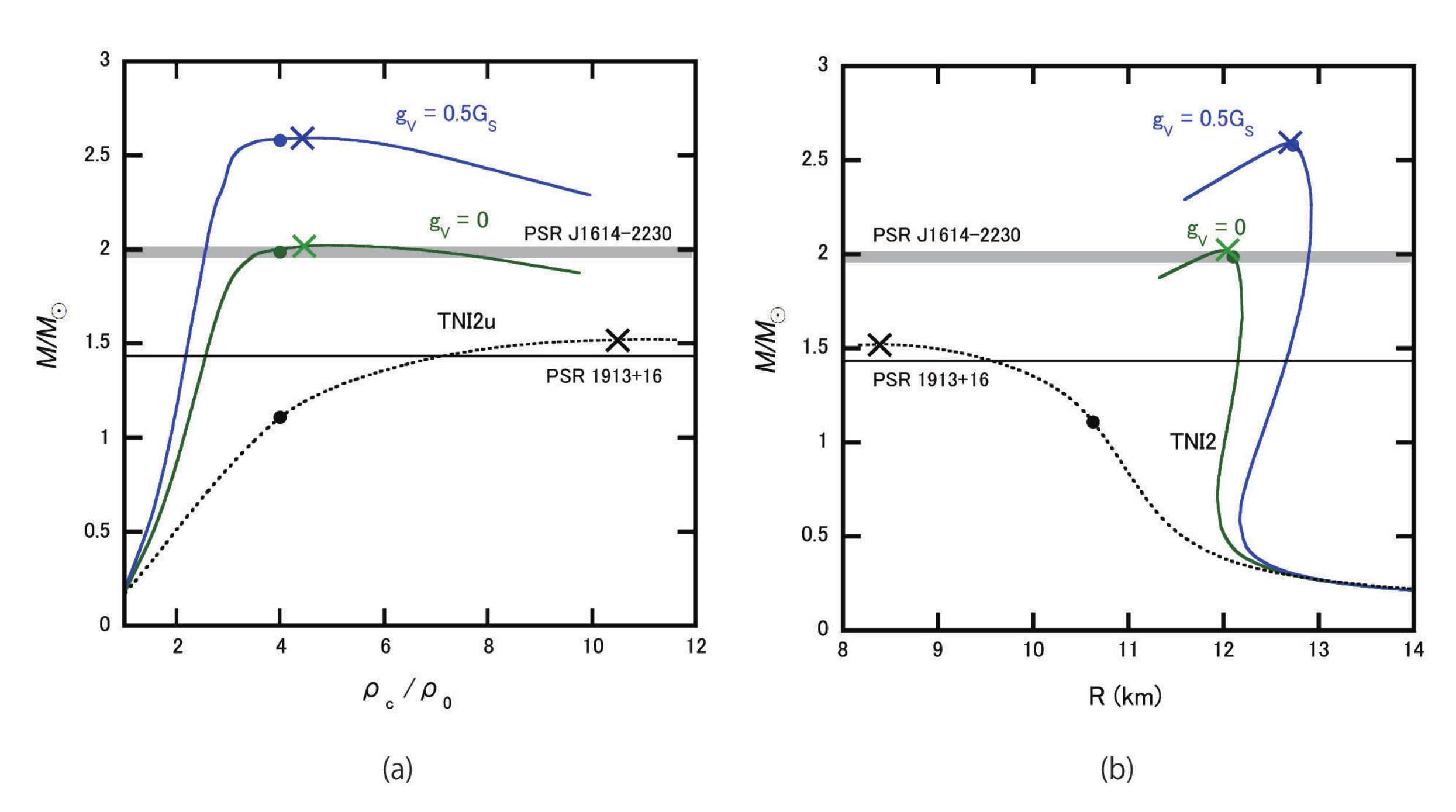}
  }
\end{center}
\caption{\footnotesize{
(a) The $M-\rho_c$ relations with the
CRover EOS (solid lines) and the TNI2u hadronic EOS (the dotted line).
 The crossover window  is fixed to be $(\bar{\rho},\Gamma)=(3\rho_0,\rho_0)$.
 The cross symbols denote the points of $M_{\rm max}$, while 
 the filled circles denote the points beyond which the strangeness appears.
 The gray band denotes $M=(1.97\pm0.04)M_{\odot}$ for PSR J1614-2230.
 The solid black line denotes $M=1.44M_{\odot}$ for PSR 1913+16.  
(b) 
  The $M-R$ relation with the same EOSs as (a).  
  These figures are adapted from \cite{Masuda:2012ed}.
 }}
      \label{m-rho-tni2u-energy}
 \end{figure*} 

In Fig.\ref{interpolate-e-tni2u-sound} (a), we show the
interpolated EOS at $T=0$ (cold CRover EOS)
with the TNI2u as a hadronic EOS and the NJL with $g_{_V}=0$ ($g_{_V}=0.5 G_{_S}$) 
as a quark EOS. The diquark condensates are switched off here ($H=G_D'=0$).
The sound velocity squared as a a function of $\rho$ is shown in Fig.\ref{interpolate-e-tni2u-sound} (b).  
In both figures, the onset of the strangeness is indicated by the filled circles.
From both figures, one finds that the CRover EOS becomes stiffer than the quark phase and the hadronic phase
in the crossover region indicated by the shaded band.  
Such stiffening is induced by the extra pressure originating from the derivative of $\rho$ acting 
on $w_{\pm}(\rho)$ in Eq.(\ref{eq:P-mu-K-v}):  
\begin{eqnarray}
P(\rho)={P}_{\rm H}(\rho) w_-(\rho)+{P}_{\rm Q}(\rho)  w_+ (\rho) + \Delta P (\rho).
\end{eqnarray}
The extra pressure $\Delta P$ is required from thermodynamic consistency and 
has a property, $\Delta P(\rho=0)=\Delta P(\rho=\infty)=0$ by definition,
i.e. it is a function localized in the crossover region. 
 
 By turning on CSC with $H=G_S$ and $G_D'=G_D$,
 the interpolated EOS becomes a little bit softer than the case without CSC 
 in the crossover region as shown by the red lines in 
 Fig.\ref{csc-sound} (a) and (b). Associated with this, the onset density of the strangeness 
 is reduced from 4$\rho$ to $3.6 \rho_0$.
 As we have discussed in \S 4,
 there is little room for the CFL phase to appear inside NSs in our CRover EOS,
 it is not considered in this figure.

\section{Neutron Stars with CRover EOS at $T=0$}

We study the structure of the spherically symmetric neutron stars in hydrostatic equilibrium 
by solving the Tolman-Oppenheimer-Volkov (TOV) equation;
\begin{eqnarray}
\frac{dP}{dr}&=&-\frac{G}{r^2}\left(M(r)+4\pi Pr^3\right)\left(\varepsilon +P\right)\left(1-2GM(r)/r\right)^{-1} , \nonumber \\
M(r)&=&\int^r_0 4\pi r'^2 \varepsilon(r') dr' .
\end{eqnarray}
Here $r$ being the 
radial distance from the center and $G$ is the gravitational constant.

\subsection{Case without Color Superconductivity}
 
First, we consider the case without the diquark condensate ($H=G_{_D}'=0$) as a baseline.
In Fig.\ref{m-rho-tni2u-energy} (a),
we plot $M-\rho_c$ relation from the CRover EOS (interpolation between the 
TNI2u  hadronic EOS and the NJL quark EOS with $g_{_V}/G_{_S}=0, 0.5$ 
in the crossover region $(\bar{\rho},\Gamma)=(3\rho_0,\rho_0)$).
For comparison, the $M-\rho_c$ relation only with TNI2u hadronic EOS is plotted by the dotted line.  
Fig.\ref{m-rho-tni2u-energy} (b) shows the corresponding $M-R$ relation for the same 
EOSs as Fig.\ref{m-rho-tni2u-energy} (a).
Strong correlation between the sudden stiffness of the sound velocity in Fig.\ref{interpolate-e-tni2u-sound} (b)
and the NS masses in Fig.\ref{m-rho-tni2u-energy} (a,b) can be seen.
Also, as $g_{_V}$ increases from zero, the quark EOS and hence the CRover EOS become stiffer, which 
 increases the maximum mass (indicated by the cross symbol) beyond 2$M_{\odot}$. 
 
One important aspect of the present result  is that the radii of NSs 
are confined in a narrow range: For example, in the case of $g_{_V}/G_{_S}=0 (0.5)$, 
all the NSs with $0.5 < M/M_{\odot} < 2.0$ ($0.5 < M/M_{\odot} < 2.5$)
 have the radius in the range  $R= (12.0 \pm 0.2) $ km ($R= (12.5 \pm 0.5) $ km).
Such a narrow window of  $R$ independent of the values of $M$ 
 will confront the present and future observations of the neutron star radii 
 \cite{Steiner:2012xt,Ozel:2015fia}.

  \begin{figure*}[!t]
  \begin{center}
  \resizebox{0.9\textwidth}{!}{
  \includegraphics{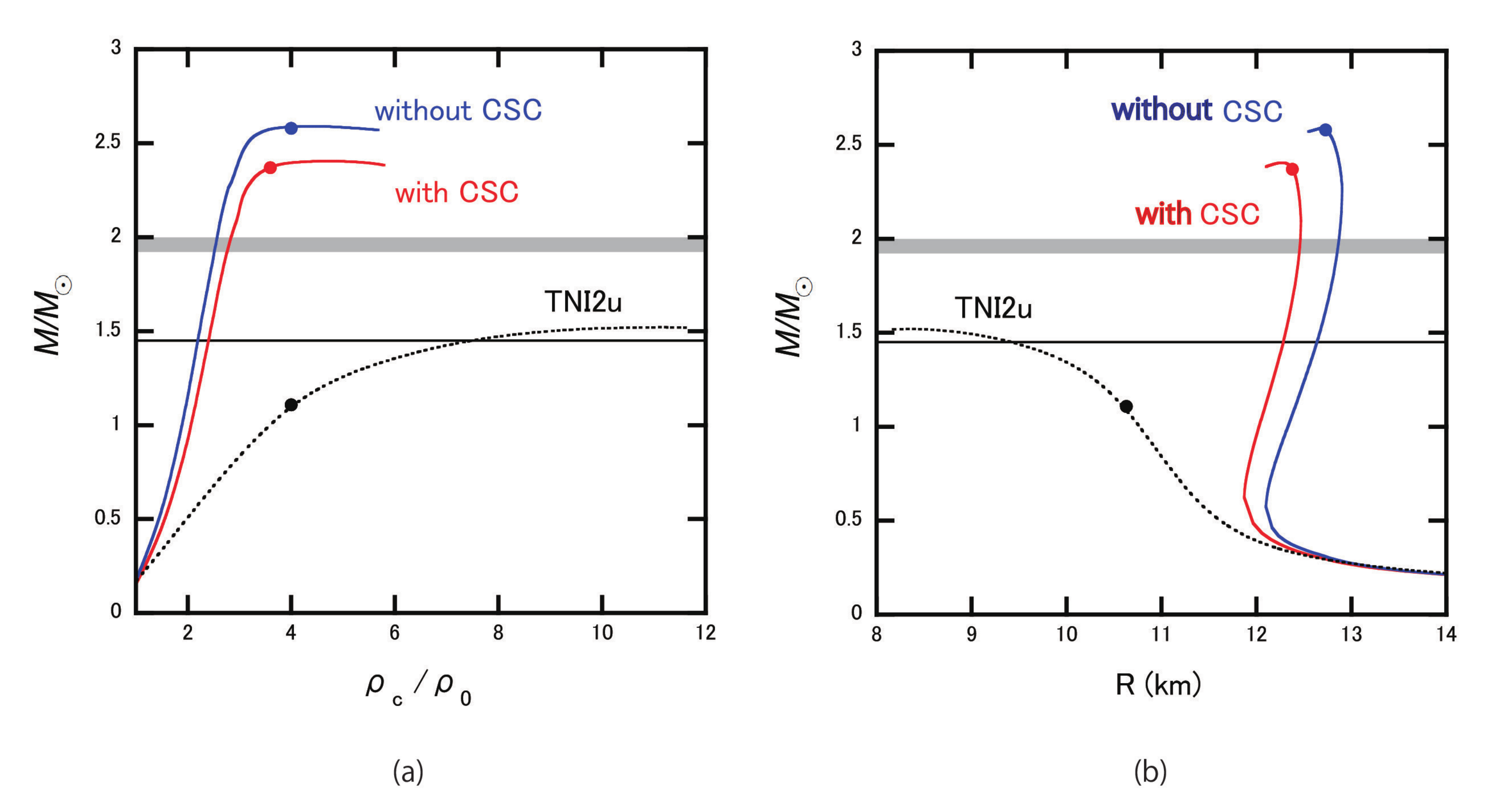}}
  \end{center}
      \caption{\footnotesize{
(a) The $M-\rho_c$ relations with the CRover EOS (solid lines) 
and TNI2u hadronic EOS (dotted line)
with and without the CSC phase for $g_{_V}/G_S = 0.5$.   
(b) 
The $M-R$ relation with the same EOSs as (a).
 }}
      \label{csc-m-rho}
 \end{figure*} 
 
 \begin{table*}[!b]
\caption{\footnotesize{
$M_{\rm max}/M_{\odot}$ ($\rho_c/\rho_0$) for 
 different choices of hadronic EOS, quark EOS and the crossover window.
 }} 
\label{table 1}
\centering
  \begin{tabular}[c]{c|c|c|c|c|c}  \hline \hline
 & \multicolumn{2}{c|}{\shortstack{$g_{_V}=0$ \\ without CSC}}
 &\multicolumn{2}{|c|}
 {\shortstack{$g_{_V}=0.5G_{_S}$ \\ without CSC}} 
& {\shortstack{$g_{_V}=0.5G_{_S}$ \\ with CSC}}\\
\cline{2-3}  \cline{3-4} \cline{4-5} \cline{5-6} 
  $(\bar{\rho},\Gamma)$& $(3\rho_0,\rho_0)$ & $(5\rho_0,2\rho_0)$ & $(3\rho_0,\rho_0)$ & $(5\rho_0,2\rho_0)$ & $(3\rho_0,\rho_0)$  \\   \hline 
 TNI2u   & 2.02 (4.5) & 1.86 (8.7) & 2.59 (4.4) & 2.25 (6.1) & 2.40 (4.9) \\ 
 TNI2    & 2.02 (5.8) & 1.84 (9.1)  & 2.59 (4.3) & 2.23 (6.8) & 2.40  (4.8)\\ 
\hline \hline
\end{tabular} 
\end{table*}

Another important aspects of the present result is the onset of the 
strangeness inside the NS core as indicated by the filled circles in 
Fig.\ref{m-rho-tni2u-energy}(b):
The strangeness appears only in very massive NSs with nearly the maximum mass
if we have hadron-quark crossover. In fact, $\rho_c=2.4 \rho_0$ for 2$M_{\odot}$ while
$\rho_c=4.4 \rho_0$ for $2.59 M_{\odot}$ under the CRover EOS with $g_{_V}=0.5G_S$,
so that even the observed 2$M_{\odot}$ NSs are unlikely to have strangeness inside.

\subsection{Case with Color Superconductivity}

In Fig. \ref{csc-m-rho} (a,b),
$M-\rho_c$ and $M-R$ relations are plotted  by using the 
CRover EOS with and without CSC given in  Fig.\ref{csc-sound}(a).
For comparison, the results of the TNI2u hadronic EOS are shown by
 the black dotted lines.
As we discussed in the last section, the CSC softens the EOS. 
Then, the $M_{\rm max}$ of the NS with CSC
becomes smaller by  0.2$M_{\odot}$  than that without CSC phase.
Such a reduction of $M_{\rm max}$ due to CSC is consistent with 
other calculations (see e.g.,  \cite{Alford:2004pf,Lastowiecki:2015mpa}).

Two remarks are in order here about the effect of CSC on the $M-R$ relation:
(i) The central density of the NSs does not exceed 4.9$\rho_0$ in CRover EOS with 
CSC, so that the CFL phase barely appears inside the star. 
(ii) The effect of 2SC already becomes visible
 for low mass stars ($M< 0.5 M_{\odot}$) with the central density below 2 $\rho_0$. 
This is because we have a smooth interpolation between the hadronic EOS and quark EOS, so that
the 2SC component has small but non-negligible contribution even below 2 $\rho_0$. 
Physically,  this  could be interpreted as partial percolation of the nucleons
into quarks with strong diquark correlations.

Table \ref{table 1} is a summary of the the maximum mass  $M_{\rm max}$ and 
the central density $\rho_c$ normalized by  $\rho_0$ with the CRover EOS.
Two sets of hadronic EOS (TNI2u and TNI2) are adopted, but the difference is small. 
The strength of the repulsive vector interaction $g_{_V}$  and the 
crossover density $\bar{\rho}$
are changed to see the sensitivity of the results.
As $g_{_V}$ becomes larger and
the $\bar{\rho}$ becomes smaller,
the maximum mass $M_{\rm max}$ increases due to the presence
of the quark matter.  The effect of CSC generally decreases $M_{\rm max}$  and
increases $\rho_c$.  As long as $\bar{\rho}$ is around 3$\rho_0$, 
the CRover EOS can easily accommodate the 2$M_{\odot}$ NSs.

\begin{figure*}[!t]
\centering
\resizebox{0.5\textwidth}{!}{
  \includegraphics{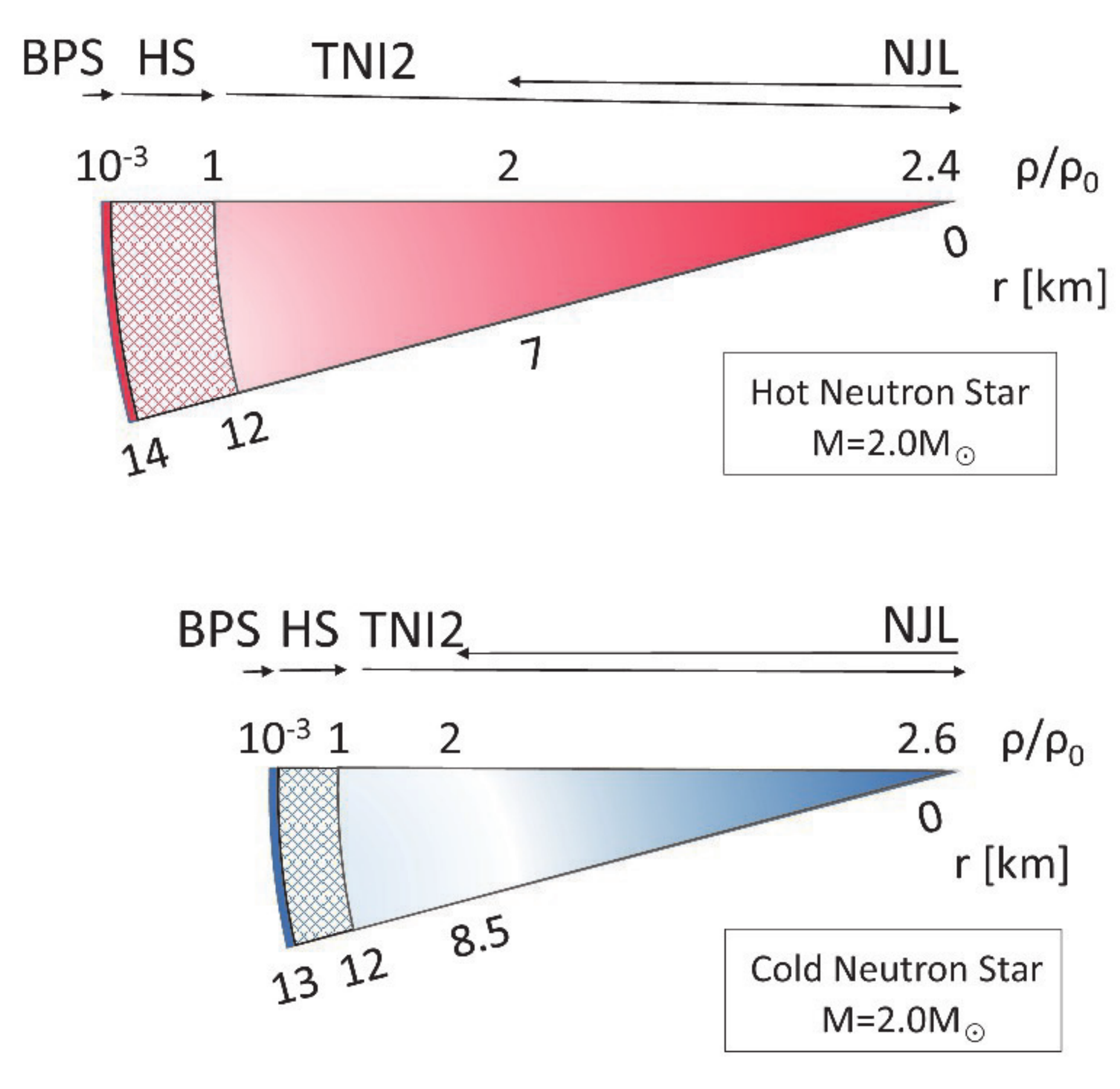} }
\caption{\footnotesize{
 Comparison between the
 hot NS and cold NS  with $M=2M_{\odot}$ obtained by the CRover EOS with $g_{_V}=0.5G_S$.
 We take $(Y_l, \hat{S})=(0.3, 1)$ to characterize the hot NS.
 As for the details of the  EOSs adopted at different densities, BPS, HS, TNI2 and NJL, see the text.
 This figure is adapted from \cite{Masuda2015}.
} }
    \label{fig:hotschematic}
\end{figure*} 

\section{Neutron Star with CRover EOS at $T\neq 0$}

In this section, we describe the properties of hot neutron stars  created after the
 core-collapsed Type-II supernova explosion by generalizing the idea of the hadron-quark crossover
 discussed in the previous sections
 (For detailed, see \cite{Masuda2015}). 

 During the first few seconds after the core bounce, 
 the proto-neutron star with the radius $\sim$ 200 km is formed.
 It undergoes a rapid contraction and  evolves into  a ``hot'' neutron star
 with the radius $\sim$ 20 km in about 1 second (or to a black hole).
 The hot NS at birth in quasi-hydrostatic equilibrium is characterized by the supernova matter 
 with the lepton fraction, $Y_l=Y_e+Y_{\nu}\sim 0.3-0.4$, and 
 the entropy per baryon, $S/N=\hat{S}\sim 1-2$.
 The neutrinos are trapped inside the hot NSs
 and contribute to the $\beta$-equilibrium.
 With this as an initial condition, the hot NS contracts gradually by 
 the neutrino diffusion  with the time scale of 10-100 seconds and evolves to a 
 nearly ``cold'' NS with $Y_{\nu}\simeq 0$ and $\hat{S}\simeq 0$,
 unless another collapse to a black hole does not take place \cite{Prakash:2000jr,Janka:2012wk,Roberts:2012zza}.

 Thermal properties of the hot NSs are intimately related to the physics of 
 high density matter at finite temperature. Indeed, 
 the hot NSs with the hadron-quark mixed phase (Fig. \ref{crossover-1}(a))
 have been studied previously, 
 e.g. \cite{Prakash:1995uw,Nakazato:2008su,pagliara,Chen:2013tfa}. 
 It is then interesting to explore properties of the hot NSs (such as the $M-R$ relation
and the profiles of the temperature, density and sound velocity inside the star etc)
 with hadron-quark crossover  (Fig. \ref{crossover-1}(b)).
   
 In Fig.\ref{fig:hotschematic}, we show a schematic picture which compares the 
 internal structure of the hot and cold NSs with $2M_{\odot}$. 
 Above the normal nuclear matter density $\rho_0$, we use the 
 EOS interpolated between TNI2 (hadron) and NJL (quark). 
 On the other hand, below $\rho_0$,
 we use the thermal EOS which  consists of an ensemble of nuclei
 and interacting nucleons in nuclear statistical equilibrium given 
 by Hempel and Schaffner-Bielich  (HS EOS) \cite{Hempel:2009mc}.
 (Use of other EOSs below $\rho_0$ does not show quantitative difference 
 as discussed in \cite{Buyukcizmeci:2012it}. )
 Once the baryon density becomes smaller than the neutron drip density $10^{-3} \rho_0$, 
 the temperature becomes smaller than 0.1 MeV.
 Then we switch to the standard BPS EOS \cite{Baym:1971pw}.

\subsection{Supernova Matter at finite $T$}

 \begin{figure*}[!t]
\centering
\resizebox{0.86\textwidth}{!}{
  \includegraphics{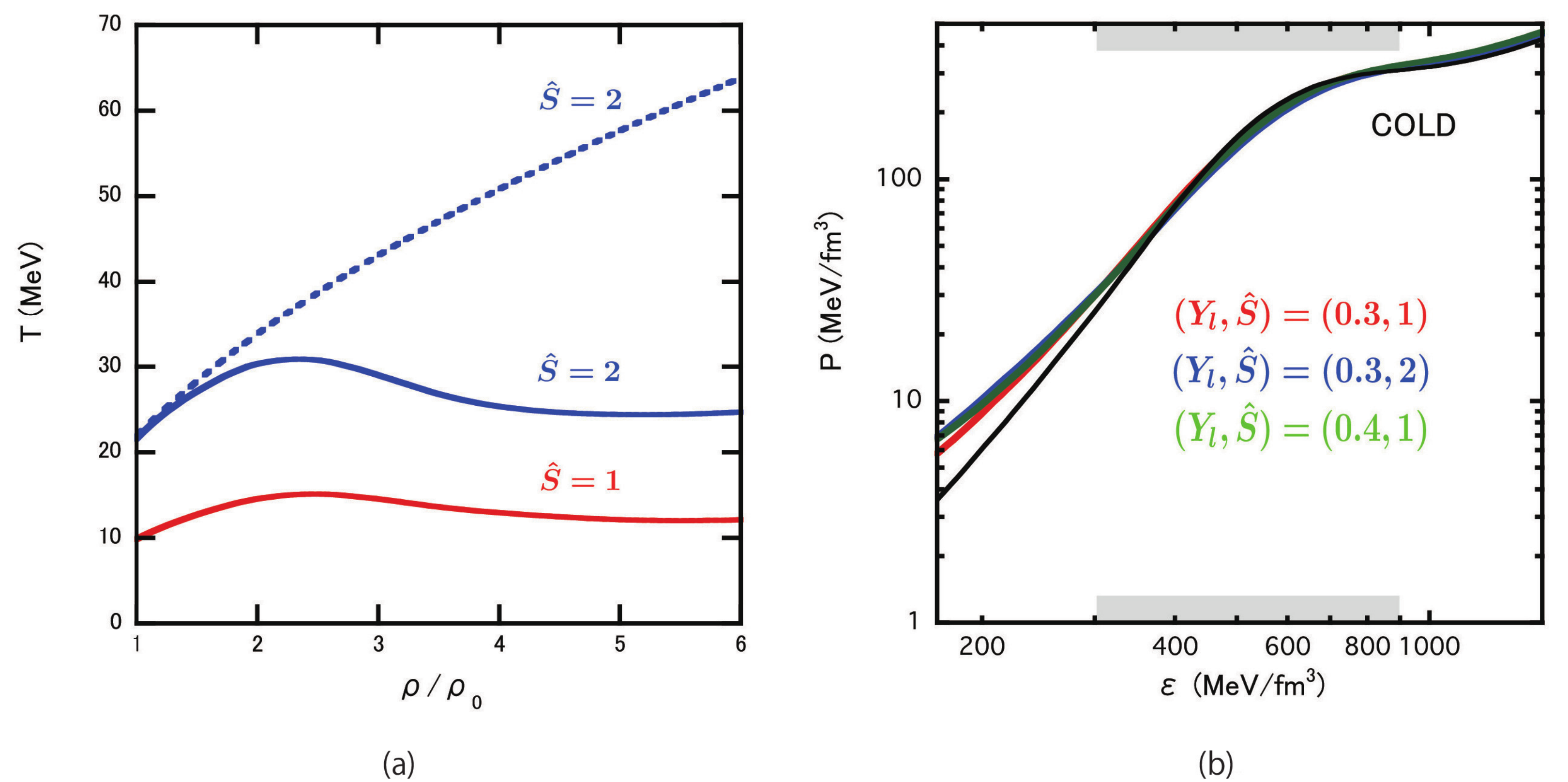}
  }
\caption{\footnotesize{
(a) The temperature $T$ of the isentropic matter as a function of the baryon density $\rho$
for $Y_l=0.3$ with $\hat{S}=1$ (red) and $\hat{S}=2$ (blue) in hot CRover EOS.
 The dashed line corresponds to the hot TNI2 EOS only with hadrons and leptons.
(b) The isentropic pressure $P$ of hot CRover EOS as a function of $\varepsilon$ for
$(Y_l, \hat{S})=$(0.3, 1), (0.3, 2) and (0.4, 1).
The black line corresponds to the cold CRover EOS for cold neutron star matter.
The crossover window  is shown by the shaded area on the horizontal axis. 
These figures are adapted from \cite{Masuda2015}.
  } }
    \label{fig:crossoverentropy}
\end{figure*}  

\begin{figure*}[!t]
\centering
\resizebox{0.86\textwidth}{!}{
  \includegraphics{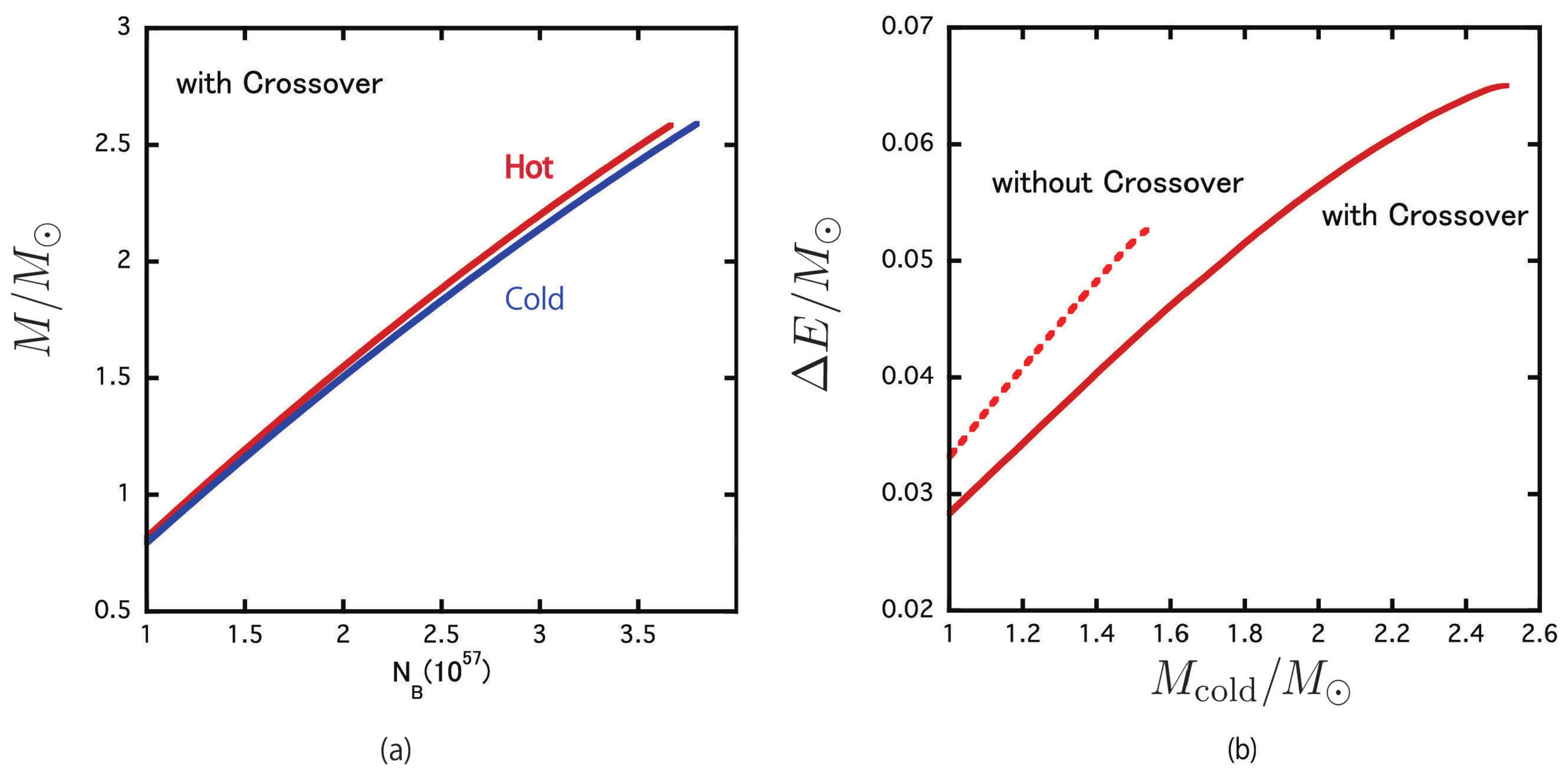}
  }
\caption{\footnotesize{
(a) The neutron star mass $M$ as a function of the total baryon number $N_B$ in CRover EOS.
Red and blue curves correspond to the hot and cold neutron stars, respectively.
(b) The energy release  $\Delta E=M_{\rm hot} - M_{\rm cold}$
 as a function of the cold neutron star mass $M_{\rm cold}$.
 $(Y_l, \hat{S})=(0.3, 1)$ is adopted.
 These figures are adapted from \cite{Masuda2015}.
 } }
    \label{fig:hot1}
\end{figure*} 
   
Major differences of the  supernova matter from that
of the  cold matter discussed in previous sections are (i) the diffused Fermi surface due to
the effect of $T$, (ii) the existence of degenerate neutrinos, and  (iii) the 
contributions from anti-particles.  

By neglecting the strangeness in hadronic matter and the muon which are irrelevant for stiff EOS,
we consider a system composed of
$n$, $p$, $e^-$, $e^+$, $\nu_e$ and $\bar{\nu}_e$ in the hadronic EOS at finite $T$, and
 $u$, $d$,$s$, $e^-$, $e^+$, $\nu_e$ and $\bar{\nu}_e$ in the quark EOS at finite $T$.
 The charge neutrality, chemical equilibrium and baryon and lepton number conservations
 are imposed. In practice, we find $\mu_e/T > 15$ in the interior of the hot NSs, so that
   the effects of $e^+$ and  $\bar{\nu}_{e}$ (as well as neutrinos in 
   second and third generations) are negligibly small.
  The color superconductivity is switched off for simplicity. 
 
We use  the Helmholtz free-energy per baryon $\hat{F}=F/N_{_B}=\hat{E}-T\hat{S}$ 
 as a basic quantity to interpolate the hadronic matter and the quark matter at finite $T$ \cite{Masuda2015}.
 This is  a natural generalization of $\hat{E}$ at $T=0$ in the previous sections.
 $\hat{F}$   is a function of  $\rho$, $T$ and $Y_l$, so that we have
\begin{eqnarray}
\hat{F}(\rho,T; Y_l)&=&\hat{F}_{\rm H}(\rho,T; Y_l) w_-(\rho,T) \nonumber \\
&&+\hat{F}_{\rm Q}(\rho,T; Y_l)  w_+ (\rho,T).
\label{eq:new-HQ-EOS}
\end{eqnarray}
Here $\hat{F}_{\rm H}$ and $\hat{F}_{\rm Q}$ are
 the hadron+lepton free-energy per baryon and the quark+lepton  free-energy per baryon, respectively.
The typical temperature of hot NSs is about $30$MeV which is 
sufficiently smaller than the thermal dissociation temperature of hadrons 
 of about $200$ MeV.  Therefore, we ignore the $T$-dependence of the weight functions,
 $w_{\pm}(\rho,T) \rightarrow w_{\pm}(\rho)$. 

\begin{figure*}[!t]
\centering
\resizebox{0.9\textwidth}{!}{
  \includegraphics{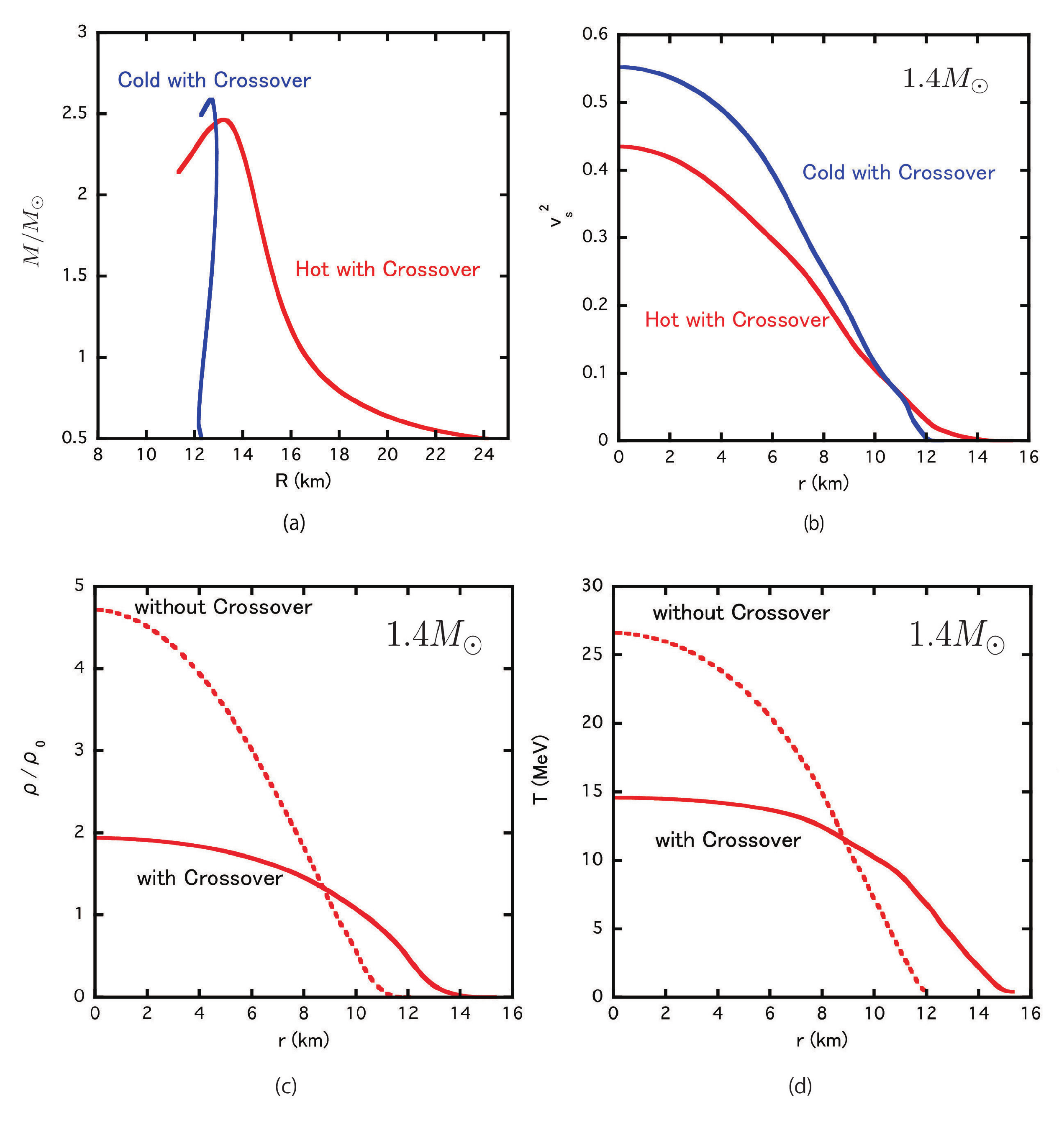}
  }
\caption{\footnotesize{
(a) Mass-Radius ($M$-$R$) relationship for $(Y_l, \hat{S})=(0.3, 1)$.
Red: hot neutron stars with hot CRover EOS.
Blue: cold neutron stars with cold CRover EOS. 
(b) The sound velocity squared $v_s^2$ as a function of the distance from the center $r$
of $1.4M_{\odot}$ neutron star.  Colors on each line are the same as in (a).
(c) The density profiles of the hot neutron star with $M=1.4M_{\odot}$ and $(Y_l, \hat{S})=(0.3, 1)$.
Solid and dashed lines correspond to the EOS with crossover and without crossover, respectively.
(d) The temperature profiles of the same neutron star as plotted in (c).
These figures are adapted from \cite{Masuda2015}.
 } }
    \label{fig:hot2}
\end{figure*}  

The entropy per baryon and the energy per baryon  are obtained by using the thermodynamic relations,  
$\hat{S}=-\partial \hat{F}/\partial T$ and $\hat{E}=\hat{F}+T\hat{S}$.
Under the assumption that $w_{\pm}$ is $T$-independent, Eq.(\ref{eq:new-HQ-EOS}) leads to
\begin{eqnarray}
\hat{S}(\rho,T; Y_l)&=&\hat{S}_{\rm H}(\rho,T; Y_l) w_-(\rho)+\hat{S}_{\rm Q}(\rho,T; Y_l)  w_+ (\rho), \nonumber
\label{cross-S} \\
\hat{E}(\rho,T; Y_l)&=&\hat{E}_{\rm H}(\rho,T; Y_l)w_-(\rho)+ \hat{E}_{\rm Q}(\rho,T; Y_l) w_+ (\rho). \nonumber
\label{energy density} 
\end{eqnarray} 

 To obtain  $\hat{F}_{\rm H}$ in Eq.(\ref{eq:new-HQ-EOS}),
  we solve the thermal Hartree-Fock equation for 
  isothermal matter composed of $n,p,e^{-},$\\
  $e^{+},\nu_e$ and $\bar{\nu}_e$
 with the same effective nucleon interaction as the TNI2 and TNI2u EOS at $T=0$
 (details are shown in \cite{Takatsuka:1994pm}). 
 To obtain $\hat{F}_{\rm Q}$ in Eq.(\ref{eq:new-HQ-EOS}),
 we start with the  Gibbs free energy calculated in the NJL model in \S\ref{sec:NJL};
 $\Omega_{\rm Q}(\mu, V, T; \mu_l)=\Omega_{\rm quark}(\mu, V, T; \mu_l) +\Omega_{\rm lepton}(\mu, V, T; \mu_l)$ 
 with $\mu$ and $\mu_l$ being  the baryon chemical potential and 
 the lepton chemical potential, respectively.  Then we make the Legendre transformation 
 from the Gibbs free energy $\Omega_{\rm Q}$ to the Helmholtz free-energy  $F_{\rm Q}$ \cite{Masuda2015}.
 
 \begin{figure*}[!t]
\centering
\resizebox{0.9\textwidth}{!}{
  \includegraphics{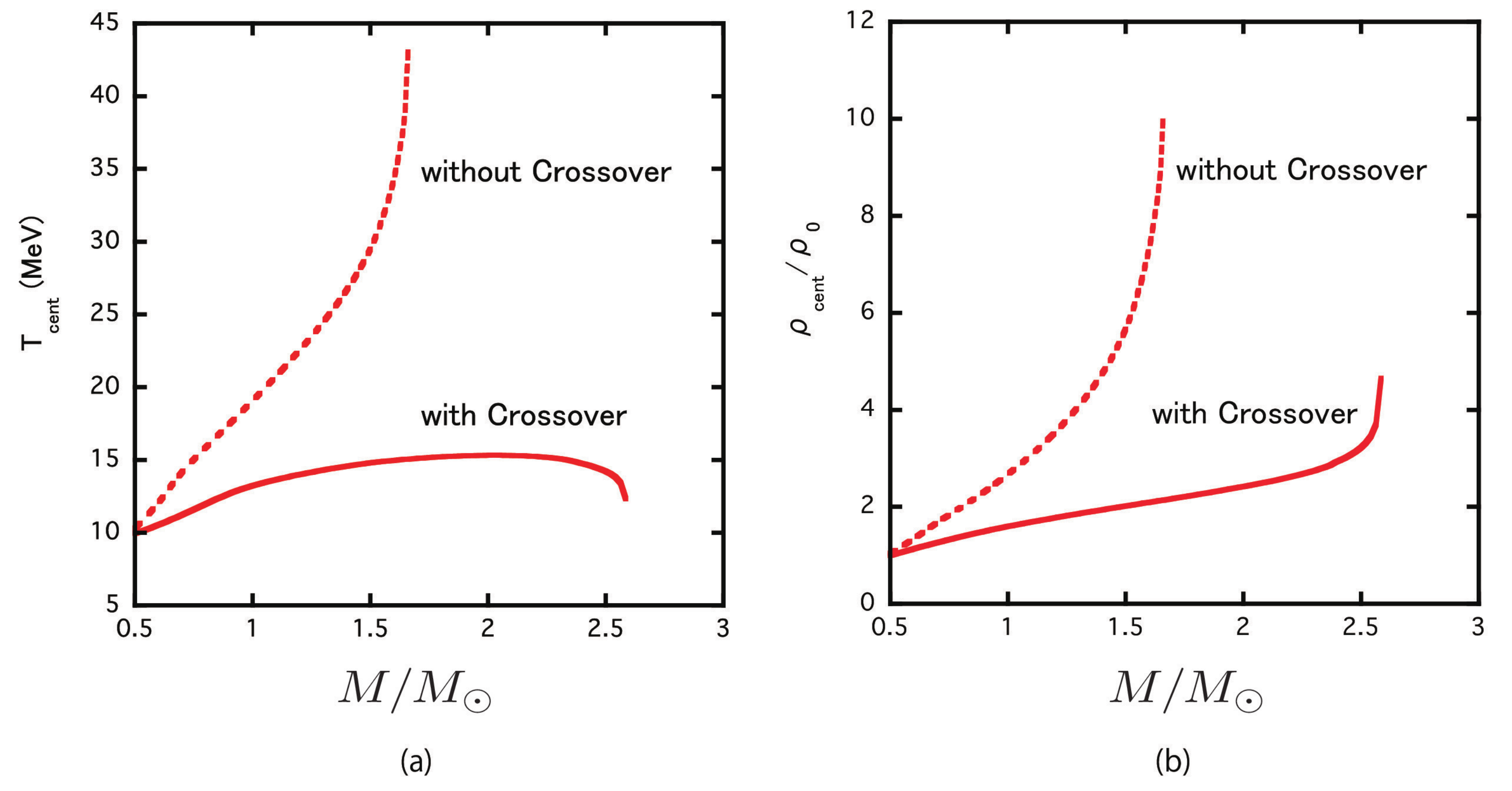}
  }
\caption{\footnotesize{
(a) The central temperature $T_c$ and (b) the central density
$\rho_{\rm cent}$ as a function of the neutron star mass $M$ of hot NS 
with $( Y_l ,\hat{S} )$=(0.3,1). The solid (dashed) lines 
correspond to the EOS with (without) crossover.
These figures are adapted from \cite{Masuda2015}.
 }}
    \label{fig:hot3}
\end{figure*}   
 
In the following,  we consider typical values of  
the lepton fraction $Y_l=0.3, 0,4$ and those of the entropy per baryon $\hat{S}=1, 2$. 
 The crossover window is fixed to be  $(\bar{\rho},\Gamma)=(3\rho_0,\rho_0)$, 
 and $g_{_V}=0.5 G_s$ is adopted.

The thermodynamic quantities for isothermal matter with $\rho$, $T$ and $Y_l$ as parameters
can be converted into those for isentropic matter with $\rho$, $S$ and $Y_l$ as parameters
 by using the relation $T=T(\rho; S, Y_l)$.
 In Fig. \ref{fig:crossoverentropy}(a),
the temperature $T$ of the isentropic matter is shown as a function of the baryon density $\rho$
for $Y_l=0.3$ with $\hat{S}=1$ and $\hat{S}=2$.

The isentropic pressure is obtained through the thermodynamic relation
 \begin{eqnarray}
 P(\rho,T(\rho; S, Y_l);Y_l, \hat{S})
 =  \rho^2 \left. \frac{\partial \hat{E}}{\partial \rho} \right|_{\hat{S},N,Y_l} .
 \end{eqnarray}
The hot CRover EOS for supernova matter  is shown in Fig. \ref{fig:crossoverentropy}(b)
for $(Y_l,\hat{S})=(0.3, 1), (0.3, 2)$ and $(0.4, 1)$.
For comparison, 
the cold CRover EOS for cold neutron star matter ($T=0$ without neutrino degeneracy)
is shown by the black solid lines. The 
hot CRover EOS and cold CRover EOS
are similar except for the low density region where the 
hot EOS becomes stiffer.

\subsection{Properties of Hot NSs}

Fig. \ref{fig:hot1} (a) shows the gravitational mass $M$ as a function of the total baryon number $N_B$
for hot (red line) and cold (blue line)  neutron stars.
The hot neutron stars have larger mass than the  cold ones for given $N_B$.
We note that the maximum value of $N_B$ for hot NSs is smaller than that for cold NSs.
This implies that hot NSs do not have a chance of delayed collapse into black holes after the cooling.
This is in contrast to the previous works with the exotic phases such as the  pion condensation
\cite{PNS:takatsuka} and the hadron-quark mixed phase \cite{Prakash:1995uw,pagliara};
in those cases,  the softening of the EOS due to exotic phases is tamed by the finite temperature effect, 
so that the maximum value of $N_B$ for hot NS becomes larger than that of the cold NS
and there arises a chance of the delayed collapse.

 In Fig. \ref{fig:hot1}(b), we show $\Delta E=M_{\rm hot} - M_{\rm cold}$
 in the unit of $M_{\odot}$  as a function of the mass of  cold NS, $M_{\rm cold}$.
 The typical amount of energy released due to the contraction
 is about $0.04M_{\odot}$ for $M_{\rm cold}=1.4 M_{\odot}$.  

In Fig. \ref{fig:hot2} (a),
we plot $M-R$ relation for hot and cold NSs with and without crossover.
The maximum mass of hot NSs is very similar with that of cold NSs.
On the other hand, a big difference of the radius can be seen between hot and cold NSs.
This comes from the stiffening of the hot EOS  at densities below $\rho_0$.
The local sound velocity squared $v_s^2(r)$ for isentropic matter can be calculated as
\begin{eqnarray}
v_s^2 (\rho;Y_l, \hat{S})= \left. \frac{\partial P}{\partial \varepsilon} \right|_{Y_l,\hat{S}}
=  \left. \frac{d P(\rho,T(\rho); Y_l, \hat{S})/d\rho}{d \varepsilon(\rho,T(\rho); Y_l, \hat{S})/d\rho} \right|_{Y_l,\hat{S}}
\end{eqnarray} 
with $\rho(r)$ obtained by the TOV equation. 
In Fig.\ref{fig:hot2} (b), sound velocity squared is  plotted
as a function of the distance from the center $r$ for $M=1.4M_{\odot}$.
The sound velocity in cold NS is larger (smaller) at higher (lower)  density than that of the hot NS \cite{Masuda2015}.

To see the thermal and lepton effects on the internal structure of the hot NSs,
we plot  the density profile $\rho(r)$ and the temperature profile $T(r)$ of hot NSs with canonical $1.4M_{\odot}$ in Fig. \ref{fig:hot2} (c) and (d), respectively.
Due to the stiffness of the CRover EOS,
the central density becomes smaller and the profile becomes flatter as shown in Fig. \ref{fig:hot2}(c).
Moreover,
as we have shown in Fig.\ref{fig:crossoverentropy}(a), 
$T$ becomes smaller for given $\rho$ by the crossover to quark degrees of freedom.
Those are the reasons why the temperature of the internal core of the hot NS becomes
smaller and flatter with crossover (Fig. \ref{fig:hot2}(d)).

In Fig. \ref{fig:hot3} (a) and (b),
the  central temperature $T_{\rm cent}$ and  the central density $\rho_{\rm cent}$
of the hot neutron stars with and without the crossover are plotted  
as a function of neutron star mass $M$. The effects of crossover on the internal structure of the 
NSs shown in Fig. \ref{fig:hot2}(c,d) for $M=1.4M_{\odot}$ can be seen for 
wide range of $M$.

\section{Summary and Concluding Remarks}

 In this article,
 we have studied the bulk properties of cold and hot neutron stars on the basis of hadron-quark crossover picture.
 A new EOS,  ``CRover", introduced in \cite{Masuda:2012kf,Masuda:2012ed,Masuda2015} 
 describes the smooth transition from hadronic matter to quark matter  in a phenomenological way. 
 The hadron-quark crossover which makes the EOS stiffer by the effect of the quark matter
 is in contrast to the first-order hadron-quark transition leading to the softening of the EOS.

 At zero temperature, a  crossover at around 3$\rho_0$ 
 leads to an EOS hard enough to sustain the  $2M_{\odot}$ NSs.
 The radii of the NSs are located in a narrow region $(12.5 \pm 0.5)$ km which is 
 insensitive to their masses.
 Due to the stiffening of the EOS induced by the crossover, 
 the central density of the NSs is at most 4 $\rho_0$.
 Therefore, hyperon mixing, whose threshold density is about 4 $\rho_0$ in the CRover EOS, 
 takes place only for very massive NSs.  
This constitutes a solution of the long-standing hyperon puzzle.
  
 We have studied the effect of color superconductivity (CSC)
 on the bulk properties of NSs under the hadron-quark 
 crossover.  With the diquark coupling $H/G_{_S}=1$, we find that 2SC phase  may appear 
 inside the NSs while the onset of the CFL phase is too high to be realized even in massive NSs. 
 As a result of a slight softening of the EOS due to the color superconductivity,  
 the maximum mass of the NSs with CSC is reduced by about 0.2 $M_{\odot}$ from that without CSC.
 
 To examine the properties of the hot NSs at birth at finite temperature, we considered the supernova matter with the CRover EOS generalized to isentropic environment.
 We found that the hadron-quark crossover plays an 
 important role   to lower the central temperature of hot neutron stars 
 in comparison to the case of hadronic EOS.
  This suppression of temperature comes from the presence of the quark degrees of freedom
  in the crossover region.
  Given baryon number, hot neutron stars have generally larger radius and larger gravitational mass caused by
  the high lepton fraction and the thermal effect. This suggests that, during the contraction
  from hot NS to cold NS, gravitational energy is released and simultaneously the spin-up takes place.
  The released energy is shown to be about 0.04 $M_{\odot}$ and the spin-up rate
  is about 14 \% (assuming the conservation of angular momentum) 
  for  $M_{\rm cold}=1.4M_{\odot}$ of evolved cold neutron stars.
 
 The hadron-quark crossover turns out to have interesting phenomenological 
 implications to  the key issues of the 
 neutron stars, such as the massive neutron stars and hyperon puzzle, 
 the universal radius of the neutron stars, 
 temperature and density profiles
 inside the hot neutron stars, and so on.  One of the most important and 
 yet challenging theoretical problems is to elucidate the QCD basis 
 of the phenomenological hadron-quark 
 crossover introduced in this article.

\section*{Acknowledgment}
We thank Wolfram Weise, Gordon Baym, David Blaschke and Mark Alford for helpful discussions.
T.T. thanks 
Toshitaka Tatsumi, Shigeru Nishizaki and
especially Ryozo Tamagai who has passed away in Jan.11, 2015 for valuable discussions and interests in this work.  
K.M. thanks Mark Alford for the kind hospitality and discussions in Washington Univ. in St. Louis
where part of this work was carried out under the support of ALPS Program, Univ. of Tokyo.
K.M. is also supported by JSPS Research Fellowship for Young Scientists.
 T.H. and T.T. were partially supported by JSPS Grant-in-Aid for Scientific Research, No.25287066.
 This work was partially supported by RIKEN iTHES Project.

\end{document}